\def\UseNDSS{}{} 

\documentclass[conference]{conf_styles/ndss/IEEEtran}
\newcommand{\loadindex}[1]{}
\newcommand{\loadtail}[1]{
    \bibliographystyle{plain}
    \bibliography{#1}
}
\newcommand{\loadappend}{\appendices}

\usepackage{tikz}

\usepackage{amsmath,amssymb,amsfonts}
\usepackage{mathrsfs}

\usepackage[ruled,linesnumbered]{algorithm2e}
\SetAlFnt{\normalsize\fontfamily{cmtt}\selectfont}
\usepackage{graphicx}
\usepackage{subfig}
\usepackage{textcomp}
\usepackage{xurl}
\usepackage{xspace}
\usepackage{xcolor}
\usepackage{bm}
\usepackage{booktabs}
\usepackage{multirow}
\usepackage{amssymb}
\usepackage{pifont}
\usepackage{enumitem}
\usepackage{soul}
\usepackage[most]{tcolorbox}
\usepackage[para,online,flushleft]{threeparttable}
\usepackage{amsthm}
\usepackage[T1]{fontenc}
\usepackage{fontawesome}
\usepackage{balance}
\usepackage{marvosym}
\usepackage[available,functional,reproduced]{ndssbadges}

\newcommand \ESPClickBuy{\href{https://octopart.com/esp32-s2-devkitm-1-espressif+systems-112268552}{ESP32S2}}
\newcommand \STMFFourClickBuy{\href{https://estore.st.com/en/products/evaluation-tools/product-evaluation-tools/mcu-mpu-eval-tools/stm32-mcu-mpu-eval-tools/stm32-discovery-kits/stm32f4discovery.html}{STM32F429}}
\newcommand \STMFOneClickBuy{\href{https://stm32-base.org/boards/STM32F103C8T6-Blue-Pill}{STM32F103}}
\newcommand \ReNodeClick{\href{https://renode.io/}{Renode}\xspace}
\usepackage{etoolbox}
\makeatletter
\patchcmd{\@makecaption}
  {\scshape}
  {}
  {}
  {}
\makeatletter
\patchcmd{\@makecaption}
  {\\}
  {.\ }
  {}
  {}
\makeatother
\def\tablename{Table}

\usepackage{hyperref}
\hypersetup{
    colorlinks,
    linkcolor={red!80!black},
    citecolor={green!80!black},
    urlcolor={blue!80!black}
}

\newcommand \todo[1]{}
\newcommand \red[1]{\textcolor{black}{#1}}
\newcommand \modify[1]{\textcolor{black}{#1}}
\definecolor{mytablecolor}{rgb}{0, 0, 0}

\newcommand \ourwork{MCU-Token\xspace}

\newcommand \bsub[1]{\vspace{1pt}\noindent\textbf{#1}}

\newcommand \usednumber{$usedNum$\xspace}
\newcommand \totalnumber{$totalNum$\xspace}
\newcommand \acceptnumber{$acceptNum$\xspace}

\newcommand\blfootnote[1]{%
\begingroup
\renewcommand\thefootnote{}\footnote{#1}%
\addtocounter{footnote}{-1}%
\endgroup
}

\makeatletter
\renewcommand\footnoterule{%
  \kern-3\p@
  \hrule\@width1.0\columnwidth
  \kern2.6\p@}
  \makeatother

\newcommand \emptycirc{\faCircleO}
\newcommand \halfcirc{\faAdjust}
\newcommand \fullcirc{\faCircle}

\begin{document}

\title{From Hardware Fingerprint to Access Token: Enhancing the Authentication on IoT Devices}

\author{
\IEEEauthorblockN{
    Yue Xiao\IEEEauthorrefmark{2}\IEEEauthorrefmark{1},
    Yi He\IEEEauthorrefmark{3}\IEEEauthorrefmark{1},
    Xiaoli Zhang\IEEEauthorrefmark{4}, 
    Qian Wang\IEEEauthorrefmark{2}\textsuperscript{\Letter},
    Renjie Xie\IEEEauthorrefmark{3},
    Kun Sun\IEEEauthorrefmark{5},
    Ke Xu\IEEEauthorrefmark{3}, and
    Qi Li\IEEEauthorrefmark{3}\textsuperscript{\Letter}
}
\IEEEauthorblockA{
    \IEEEauthorrefmark{2} Key Laboratory of Aerospace Information Security and Trusted Computing, Ministry of Education, \\ School of Cyber Science and Engineering, Wuhan University, \\
    \IEEEauthorrefmark{3}Tsinghua University, 
    \IEEEauthorrefmark{4}Zhejiang University of Technology, \IEEEauthorrefmark{5}George Mason University, \\
    \IEEEauthorrefmark{2}\{yuexiao, qianwang\}@whu.edu.cn,
    \IEEEauthorrefmark{3}\{heyi21, xrj21\}@mails.tsinghua.edu.cn, \\
    \IEEEauthorrefmark{3}\{xuke, qli01\}@tsinghua.edu.cn, \IEEEauthorrefmark{4}xiaoli.z@outlook.com, \IEEEauthorrefmark{5}ksun3@gmu.edu
}
}

\ifdefined\UseCCS
    \begin{abstract}

The proliferation of consumer IoT products in our daily lives has raised the need for secure device authentication and access control.
Unfortunately, these resource-constrained devices typically use token-based authentication, which is vulnerable to token compromise attacks that allow attackers to impersonate the devices and perform malicious operations by stealing the access token.
Using hardware fingerprints to secure their authentication is a promising way to mitigate these threats.
However, once attackers have stolen some hardware fingerprints (e.g., via MitM attacks), they can bypass the hardware authentication by training a machine learning model to mimic fingerprints or by reusing these fingerprints to craft forged requests.\blfootnote{\IEEEauthorrefmark{1} The first two authors contributed equally to this paper.}\blfootnote{\textsuperscript{\Letter} Qian Wang and Qi Li are the corresponding authors.}

In this paper, we present \ourwork, a secure hardware fingerprinting framework for MCU-based IoT devices even if the cryptographic mechanisms (e.g., private keys) are compromised. \ourwork can be easily integrated into various IoT devices by simply adding a short hardware fingerprint-based token to the existing payload. To prevent the reuse of this token, we propose a message mapping approach that binds the token to a specific request by generating the hardware fingerprints based on the request payload. To defeat the machine learning attacks, we mix the valid fingerprints with poisoning data so that attackers cannot train a usable model with the leaked tokens. \ourwork can defend against adversaries who may replay, craft, and offload the requests via MitM or use both hardware (e.g., use identical devices) and software (e.g., machine learning attacks) strategies to mimic the fingerprints. The system evaluation shows that \ourwork can achieve high accuracy (over 97\%) with low overhead across various IoT devices and application scenarios. 
\end{abstract}

    \loadindex{template; formatting; pickling}
\fi

\ifdefined\UseNDSS
    \IEEEoverridecommandlockouts
    \makeatletter\def\@IEEEpubidpullup{6.3\baselineskip}\makeatother
    \IEEEpubid{\parbox[b]{\columnwidth}{
        Network and Distributed System Security (NDSS) Symposium 2024\\
        26 February - 1 March 2024, San Diego, CA, USA\\
        ISBN 1-891562-93-2\\
        https://dx.doi.org/10.14722/ndss.2024.241231\\
        www.ndss-symposium.org
    }
    \hspace{\columnsep}\makebox[\columnwidth]{}
    }
\fi

\maketitle

\ifdefined\UseCCS
\else
    
\fi


\section{Introduction}
The emerging Internet of Things (IoT) technologies have been widely applied in various areas of our daily life. For instance, passive keyless entry (PKE) systems~\cite{HODOR} can remotely unlock and activate the vehicles with a small key fob, and IoT hardware security tokens (HSTs)~\cite{u2f_key} are used to protect crypto wallets or login websites as universal two-factor (U2F) authentication. 
The cost-effective and power-efficient Microcontrollers (MCUs) are widely adopted by these IoT devices since they integrate CPU, RAM, ROM, and peripherals on a single chip. Meanwhile, the low cost and high integration also limit the hardware resources available on these devices (e.g., 256KB memory, 64-300MHz clock frequency). Also, IoT devices lack of hardware protection such as memory management unit (MMU) or trusted execution environment (TEE), rendering them less secure than mobile phones and laptops.  



It is essential to ensure that MCU-based IoT devices are securely authenticated when interacting with other devices or the cloud~\cite{clone_key, smart_lock_attacks}. However, the existing token-based authentication solutions (e.g., JSON Web Token~\cite{jwt} and rolling code~\cite{rolling_code}) suffer from various attacks due to the constrained system resources and insecure implementations~\cite{attack_telsa, iot_hazards}. 
For instance, Tesla key fobs are vulnerable to key clone attacks~\cite{tesla_key_clone} and RFID/BLE relay attacks~\cite{ble_relay_attack, telsa_car_key_clone}, which allow attackers to activate vehicles by masquerading as valid keys or relaying communication to the real owner's keys. Moreover, single-function devices (e.g., U2F hardware keys~\cite{attack_u2f} and hardware wallets~\cite{blackhat_soft_attack}) can be easily cloned~\cite{clone_key, u2f_clone} once attackers obtain the internal private keys during manufacturing, retail, or usage stages. Similarly, smart homes are at risk of token compromise attacks, which enable adversaries to impersonate legitimate devices, access user data, manipulate device status, and trigger malicious rules~\cite{dtap18, trigger_action2}. 

The root cause of attacks against these IoT devices is that they can be impersonated, e.g., by compromising the communication protocols and secrets, so that the fake devices can generate the same requests to deceive their peers. 
Although unclonable hardware authentication factors have been proposed to prevent these attacks~\cite{DeMiCPU, t2pair, time-print, safetynet, contauth, IoT-ID}, 
they 
are ineffective when they are applied to MCU-based IoT devices. 
First, 
most of the required hardware features (e.g., magnetic sensor~\cite{DeMiCPU}, NAND-Flash~\cite{time-print} and TEE~~\cite{safetynet}) are not supported by most commercial-off-the-shell (COTS) MCUs. 
Although physically unclonable functions (PUF)~\cite{puf_suervey} can produce device-specific crypto keys or fingerprints, they need extra integrity circuit (IC) manufacturing procedures to provide special circuits.

Second, it is still difficult to prevent the man in the middle (MitM) adversaries~\cite{mqtt_mitm, mitm_attack, smart_lock_attacks} that can mimic the hardware fingerprints via machine learning (ML) attacks or reuse previous fingerprints in forged requests. In particular, machine learning based attacks are the main threat to these hardware feature based solutions~\cite{puf_attack, puf_lessons, puf_pipe_dream}, where the attackers can collect the leaked fingerprints to train a ML model to mimic the hardware features and predict valid unused fingerprints. 
MitM attacks are real threats for various IoT devices including USB devices~\cite{usb-mitm} (e.g., U2F and hardware wallets~\cite{attack_hw_wallet}), short distance devices (e.g., BLE and RFID Passive Keyless Entry (PKE)~\cite{ble_relay_attack}), and WiFi and Ethernet devices (e.g., Smart Home~\cite{dtap18}). Although secure communication protocols can prevent attackers from stealing fingerprints~\cite{IoT-ID, DeMiCPU}, numerous real-world exploits indicate that it can still break these secure communications by constructing different attacks, e.g., remote exploiting~
\cite{knob, bias}, stealing hard-coded crypto keys~\cite{share_keys}, and investigating unencrypted traffic in millions of IoT devices~\cite{ iot_firmware,breakmi}. 



In this paper, we develop a new authentication system called \ourwork for MCU-based IoT devices, which generates access tokens based on the commonly supported hardware features.  
\ourwork can ensure authentication security even if the existing cryptographic keys and algorithms are compromised.
In particular, it can prevent MitM adversaries from crafting requests to reuse valid fingerprints when message integrity (i.e., signature) cannot be guaranteed. 
To achieve this, we design a one-round protocol that uses fingerprints to ensure the message's integrity and thus cannot be intercepted. 
We map the message with a nonce to a hash digest and utilize the digest bits to decide hardware fingerprinting methods and settings. \ourwork can support six different hardware features to generate different fingerprints with thousands of different configurations. Consequently, it can produce tens of thousands of different fingerprints enabling multiple unique fingerprints for each request. 
Thus, once an attacker attempts to reuse a fingerprint, our \ourwork backend authentication service can easily detect the attack by identifying a mismatch between the messages and fingerprints.

Moreover, in order to prevent attackers from obtaining fingerprints for model training even when the message confidentiality (e.g., algorithms and encryption keys) is compromised, \ourwork injects noises to fingerprints that the attacker models trained on the fingerprinted are poisoned. If an attacker uses leaked fingerprints to train his model, the poisoned data cannot be used to train the model to accurately predict new fingerprints, and the \ourwork backend authentication service can easily verify if the request is legitimate by checking if a portion of the fingerprints is valid. Thus, it can effectively defend against machine learning based attacks that cannot be throttled by existing hardware feature based defenses such as PUF~\cite{puf_lessons, puf_attack}. 
\modify{
\todo{B1}
Meanwhile, data poisoning does not affect backend authentication because it uses multiple fingerprints for authentication each time, and only a few fingerprints are poisoned, allowing the unpoisoned ones to successfully pass authentication.
}


\ourwork does not rely on the security of existing cryptography mechanisms on IoT devices considering many devices may lack hardware resources to enable strong encryption protection. 
Furthermore, \ourwork is lightweight so that it can be applied on various resource-constraint embedded devices as it has small memory and storage footprints. 
The \ourwork backend authentication service can also be easily deployed on normal IoT devices or on the cloud as it only uses simple machine learning algorithms such as RandomForest and ExtraTrees.

We prototype \ourwork by $\sim$5100 LoC including $\sim$3900 lines of C code for the client runtime and $\sim$1200 lines of Python code for the \ourwork's backend service, which is open source on Github\footnote{\href{https://github.com/IoTAccessControl/MCU-Token}{https://github.com/IoTAccessControl/MCU-Token}}. 
We conduct experiments on different real-world devices, which demonstrate that \ourwork is robust to existing attacks. Due to the difficulty of breaking our fingerprint binding mechanism, e.g., by manipulating payloads, attackers can hardly forge messages and reuse the fingerprints to bypass \ourwork. The success rate of mimicking fingerprints via hardware or software approaches is less than 1\%. 
In the meantime, \ourwork achieves an average 97.78\% true positive rate (TPR) with 4.87\% false positive rate (FPR) in identifying 60 devices with 3 different MCUs. 
Moreover, the incurred overhead is reasonable low. 
The additional power consumption is less than 4\% 
and the average extra authentication time is around 31ms.






In summary, our contributions are three-fold: 

\begin{itemize}
    \item We perform a systematic study on hardware features for fingerprinting the COTS MCUs.  
    We explore six key hardware features and theoretically analyze and experimentally verify the sources of hardware uniqueness.

    \item We propose a new hardware fingerprint based authentication mechanism called \ourwork, which utilizes a novel ML based design to protect device authentication without relying on cryptographic mechanisms. It binds fingerprints to specific requests and injects poisoned data to defeat different adaptive attacks, e.g., ML based attacks. 

    \item We prototype \ourwork and demonstrate its usability and performance by extensively evaluating it on 60 IoT devices of three types across three real-world scenarios, i.e., PKE/BLE key fobs, smart home sensors, and FIDO-U2F~\cite{fido_iot} hardware tokens. 
    
\end{itemize}



\section{Background}



\subsection{Hardware-based Authentication}

Various hardware-based authentication mechanisms~\cite{time-print, modern_iot_auth} are proposed to secure IoT authentication.
These approaches consider various hardware features such as physical signal characteristics~\cite{HODOR, RFID}, magnetic  characteristics~\cite{DeMiCPU}, sensors with human interaction~\cite{contauth, t2pair, wearable_pair}, or physically unclonable functions (PUFs)~\cite{puf-taxonomy}. 
Based on the authentication methods, these approaches can be categorized into two types. 

\bsub{Hardware Fingerprint as New Device Identifier.}
These approaches utilize hardware-derived data to generate unique device fingerprints as the device identifier to distinguish different devices. 
The fingerprint data can be uploaded along with the payload or concealed as side-channel information, like physical signal strength~\cite{HODOR} or time delay~\cite{time-print}.

%

\bsub{Hardware-Involved Challenge Response Authentication Protocol.}
Instead of directly identifying devices via static device identifiers generated by hardware features, challenge-response based approaches utilize diverse challenges as input to the hardware features, obtaining variable responses for authentication.
For instance, the arbiter PUF~\cite{puf_lessons, puf_mutual} can use different bits as input and statistic the relative delays in the paths of the circuit as responses.
These approaches collect enough challenge-response pairs (CRPs) and store them as key-value data~\cite{puf_set_defense} in the server or learned by ML models~\cite{RF-PUF} during the device enrollment phase.
In the authentication phase, the corresponding CRP is retrieved to verify if the response is coincident with a specific challenge. 
Human interactions can also be used as challenges, e.g. T2Pair~\cite{t2pair} employs users' button or twist operations as challenges and validates the received actions on the server.

\begin{figure}[t]
    \centering
    \includegraphics[width=0.48\textwidth]{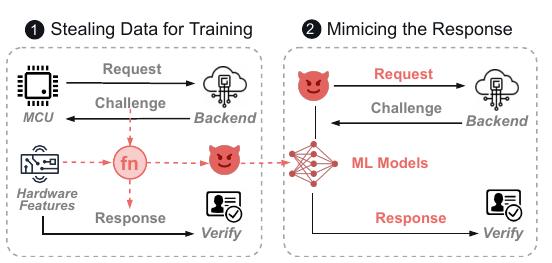}
    \caption{Machine learning attacks to PUF-based challenge-response authentication. \ding{182} Attackers eavesdrop on the communication to steal challenge-response pairs. \ding{183} Attackers learn the mapping between challenges and responses via machine learning and mimic the responses to pass the authentication.}
    \label{fig:mimic_attack_intro}
    \vspace{-15px}
\end{figure}

\subsection{Attacks to Hardware-based Authentication}
\label{sec:ml_attack}

\begin{figure}[t]
   \centering
    \subfloat[Compromised device number\label{fig:drawnapart_number}]{\includegraphics[width=0.23\textwidth]{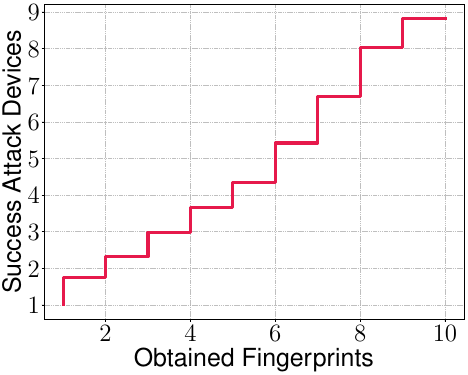}}
    \hfill 	
    \subfloat[Attack with different data size\label{fig:drawnapart_dimension}]{\includegraphics[width=0.23\textwidth]{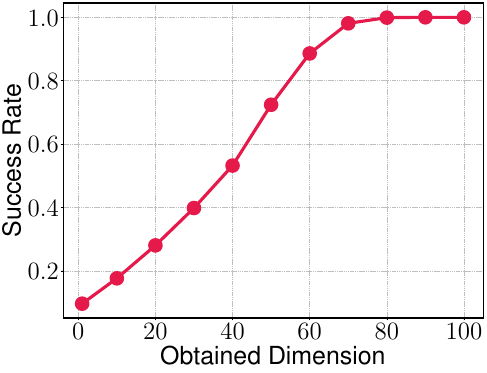}}
    \caption{Machine learning attack on IoT-ID~\cite{IoT-ID}'s ADC feature.}
    \label{fig:machine_learning_attack}
\end{figure}

The MitM adversaries should be considered in IoT scenarios, as many IoT devices are resource constraint to adopt a secure implementation of TLS with SSL pinning ~\cite{mitm_attack, iot_ssl}, or even do not encrypt the transmitted messages~\cite{breakmi}.
Under an insecure communication channel, attackers can launch two typical attacks:

\bsub{Fingerprint Mimic Attacks.}
Attackers may attempt to replicate the hardware characteristics of target devices using identical hardware or alternative devices such as FPGAs. 
This threat is addressed by existing hardware fingerprinting studies because no two devices are truly identical at IC level, and devices of the same type can still be discriminated by their micro hardware features.
However, attackers can also use software approaches (e.g.,  machine learning) to mimic the hardware features.
Figure~\ref{fig:mimic_attack_intro} shows the steps of machine learning attacks, where attackers can train a model based on a few existing fingerprints and mimic the new fingerprints.
ML attack is a common threat to both hardware fingerprints based device identifiers and the challenge-response based authentication protocols.
For instance, attackers can easily predict the responses of a given challenge for all existing PUF~\cite{puf_attack}.
The hardware features of MCU can also be easily mimicked by machine learning. 
As shown in Figure~\ref{fig:machine_learning_attack}, the features used by IoT-ID~\cite{IoT-ID} can be mimicked with high accuracy after attackers gain less than 10 unique fingerprints.

\bsub{Fingerprint Reuse Attacks.}
When the communication channel is insecure, attackers can eavesdrop and relay existing requests~\cite{attack_telsa, ble_relay_attack}, which is prevalent in existing RFID and BLE car key fobs.
MitM attackers can also replay the existing requests to reuse the fingerprint or offload the challenges of servers to real devices to get valid responses.
In this case, they do not need to mimic and generate fake fingerprints but can reuse the valid devices' real fingerprints.
They can further modify communication data while reusing the existing authentication data (e.g., hardware fingerprints, PUF response), such as changing less sensitive commands (e.g., ``turn on the light") into dangerous commands (e.g., ``open door") after compromising the encryption and signature secrets.




\subsection{Limitations of Existing Hardware-based Authentication}
As shown in Table~\ref{tab:cmp_works}, existing works including both device identifier based or challenge response based approaches are all vulnerable to fingerprint mimic attacks and fingerprint reuse attacks.
Currently, only signal-based approaches~\cite{rf_fingerprint} can resist these attacks as the physical radio features are difficult to mimic and replicate. 
However, the physical signal-based approaches can only work on wireless (e.g., RFID~\cite{HODOR} and BLE~\cite{BLE/Tracking}) devices.
This is also a common flaw of most of the existing works (except IoT-ID) that can only work on dedicated devices and are not practicable for COTS MCU.
PUF-based approaches usually require special IC fabrication processes to produce hardware discrepancies, e.g., the arbiter PUF~\cite{puf-taxonomy} requires additional arbiter circuitry and is only supported by dedicated devices (e.g., some types of NXP MCUs~\cite{nxp_puf}).


IoT-ID~\cite{IoT-ID} is the only work that supports general MCU-based IoT devices, which use commonly supported hardware features such as clock oscillators and ADC.
However, it does not take adversaries into account and the  hardware-based device identifier is just another access token that can be stolen by the attackers during transmission or at the server.
Thus, it is still vulnerable to token compromise attacks.


\begin{table}[t]
\centering
\caption{The drawbacks in existing hardware-based authentication approaches. \emptycirc, \halfcirc, and \fullcirc\xspace refer to unsupported, partial support, and full support respectively.}
\label{tab:cmp_works}
\resizebox{0.98\columnwidth}{!}{%
\begin{tabular}{@{}cccc@{}}
\toprule
\textbf{Approaches} & \textbf{\begin{tabular}[c]{@{}c@{}} Support \\ COTS MCU\end{tabular}} & \textbf{\begin{tabular}[c]{@{}c@{}} Resist Mimic\\  Attacks\end{tabular}} & \textbf{\begin{tabular}[c]{@{}c@{}}Resist Reuse\\ Attacks\end{tabular}} \\ \midrule
Signals~\cite{RFID, rf_fingerprint, HODOR, BLE/Tracking} & \halfcirc & \fullcirc & \halfcirc \\ \midrule
Human Interactions~\cite{t2pair, contauth, wearable_pair, bio_auth} & \halfcirc & \halfcirc & \emptycirc \\ \midrule
PUF~\cite{puf_mutual, puf_lessons, puf_suervey, puf-taxonomy, lightweight_puf} & \halfcirc & \halfcirc & \emptycirc \\ \midrule
Hardware Fingerprints~\cite{drawnapart_gpu,DeMiCPU,time-print,time-stamp} & \halfcirc & \halfcirc & \emptycirc \\ \midrule
\textbf{Multiple Hardware Features (\ourwork)} & \fullcirc & \fullcirc & \fullcirc \\ \bottomrule
\end{tabular}%
}
\vspace{-5px}
\end{table}

\section{Threat Model and Assumption}
\label{sec:threat_model}
We assume attackers may have compromised the communication channel and stolen the authentication tokens of valid devices.
Their attack goal is to impersonate legitimate devices to perform malicious operations, such as unlocking a car by mimicing a key fob or triggering the execution~\cite{trigger_action2} of trigger-action rules in smart homes by event spoofing~\cite{dtap18}.

As the attackers have compromised both the access control token and the encryption and signing mechanisms (if existing), they can eavesdrop and manipulate the requests of real devices or even impersonate the devices to send fake requests. To bypass the potential additional hardware-based authentication mechanisms (e.g., \ourwork), they can perform fingerprint mimic attacks to generate valid fingerprint data via software or hardware approaches, such as collecting the existing hardware fingerprint data and training their own models to mimic the hardware behaviors (software mimic attack) or using the same types of hardware to mimic the real devices (hardware mimic attack).
They can also reuse previous authentication information to send fake requests or forward the packages between the devices and the server (i.e., replay attack), or alter requests (i.e., tampering attack).

\todo{B2} 
\modify{
We assume that the collection of training data can take place in a secure environment, such as during device manufacturing in a factory, and that the training mode cannot be triggered after the training phase.
\todo{A1} 
Moreover, we assume that the device is not compromised by attackers, either locally or remotely.
Adversaries who have compromised the device are beyond our scope, and we discuss them in \S~\ref{sec:discussion}.
}

\section{System Design}

\subsection{\ourwork Overview}
\begin{figure}[t]
    \centering
    \includegraphics[width=0.48\textwidth]{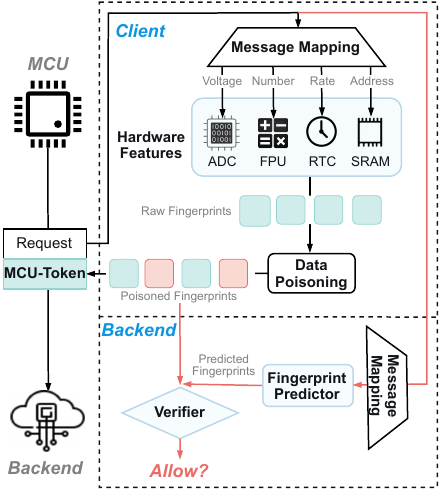}
    \caption{The architecture of \ourwork. A hardware fingerprint-based token (i.e., \ourwork) is sent along with the request from devices. The token mixes multiple valid fingerprint values (green block) with poisoned results (red block), and the backend verifies the token by comparing the fingerprints with the predicted values.}
    \label{fig:overview}
\end{figure}

Figure~\ref{fig:overview} shows the overall architecture of \ourwork. For a sensitive request, the client runtime on local devices can generate a hardware fingerprint based access token and send this token along with the requests. A client fingerprint generation module is integrated into the devices' firmware for generating an extra hardware fingerprint based access token that is sent along with the requests. A backend fingerprint verification module can be deployed on other devices or on the cloud for validating the token. 





\begin{figure}[t]
    \centering
    \includegraphics[width=0.46\textwidth]{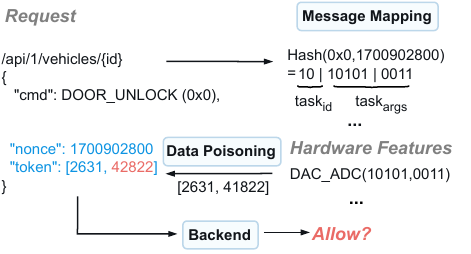}
    \caption{A running example of the Telsa car key. The blue fields are the extra payload added by \ourwork and the red text is the poisoning data.}
    \label{fig:running_example}
\end{figure}

\ourwork generates fingerprints based on the message digest of corresponding request.
Since fingerprints derived from static features may still be stolen, we use non-repetitive fingerprints for different requests.
Similar to the challenge response based approaches, our fingerprint is created by changing the fingerprint generation configurations (e.g., using different hardware features) to produce unique fingerprint values. 
To prevent attackers from impersonating the backend to send fake challenges and directly read out the devices' responses, we adopt a one-round protocol and autonomously generate the challenges from the local devices by adopting the client's message digest as the challenge code.
In this way, our fingerprint is bound to specific requests and attackers can no longer reuse fingerprints to craft requests.





\ourwork protects the fingerprints by randomly mixing poisoned results on the responses.
Our one-round protocols can still be exploited by ML attacks as our message digesting rules are known to attackers.
They can retrieve the original challenges and responses by eavesdropping on the communication and training a model to predict the responses.
It is difficult to make the hardware fingerprints unpredictable to the attackers as the responses of dedicated hardware features are simple to be fitted by machine learning.
A more practical way is to prevent attackers from obtaining enough valid fingerprints to train their model.
We choose to add poisoned data to the responses by changing some of the fingerprints to fake data, making attackers cannot distinguish the valid data and fake data.
If they train their model with the poisoned data, their model may fail to precisely predict the responses, which can be easily identified by our backend verification module.
In contrast, \ourwork is only trained with valid data in the device binding phase and does not update the model with poisoned data.
As such, \ourwork can resist machine learning attacks as our backend module is not affected by the poisoned data and can still authenticate devices based on the remaining unchanged fingerprint data.


Figure~\ref{fig:running_example} shows a running example of using \ourwork in the Telsa BLE car key. 
For sensitive commands, such as unlocking the car door, \ourwork is triggered to add a token to the request.
The token consists of two fingerprints generated by different hardware features.
Both the client and the backend implement the same message mapping algorithms that use the digest bits of the raw command and a nonce to select hardware features as fingerprinting tasks and determine the corresponding task arguments. 
One of the fingerprints is intentionally altered with poisoned data. 
The backend authentication service independently maps the task arguments from the request payload and generates two fingerprint values using the predictor.  
Access is granted when the verifier determines a close match between one of the client's fingerprints and the predicted values.





\subsection{Selecting Hardware Features}
\label{sec:choose_feature}

To ensure \ourwork can generate unique fingerprints for different requests, it is crucial to identify an adequate number of commonly supported hardware features in major MCUs.
However, previous studies~\cite{IoT-ID} only explored limited hardware features.
Therefore, we explore new hardware features by examining the datasheets of MCUs.
We identify potential hardware features by looking for theoretical evidence~\cite{puf-taxonomy} that the IC-level variation of a particular hardware feature can lead to performance or accuracy deviations. 
We then conduct experiments to validate the output variable ranges of these features across different settings or inputs, and evaluate their ability to reliably discriminate between identical devices.
\todo{C2}
\modify{Table~\ref{tab:hardware_modules}, lists some of the hardware modules on STM32F4 serials with their functional descriptions.
The features behind these modules may not have been explored or implemented on MCU devices, but their sources of uniqueness have already been revealed by existing work.
Thus, we investigate the following 6 common hardware modules of COTS MCUs:
}


\bsub{DAC/ADC.} 
A digital-to-analog converter (DAC) can convert digital values to analog signals, such as voltage. 
Conversely, an analog-to-digital converter (ADC) performs the reverse function of converting analog signals to digital outputs.
Previous studies~\cite{dac_adc_puf, IoT-ID} have demonstrated that each ADC exhibits distinct biases when outputting digital values. 
By generating multiple analog signals through the DAC, we can induce variations in the ADC's output values and use these biases to uniquely identify devices.

\bsub{Float Point Unit (FPU).}
Similar to GPU~\cite{drawnapart_gpu}, the FPU is also dedicated to accelerating float number arithmetic. 
Their computing power for float point calculations can vary among devices of various models. 
By assessing their performance in executing diverse computing tasks, we can discern and differentiate between distinct devices.

\bsub{Pulse Width Modulation (PWM).}
PWM regulates power levels by turning signals on and off at a constant frequency. 
By analyzing the accumulated power over specific time intervals at different frequencies, it is possible to differentiate between different MCUs based on the observed accumulation discrepancy.

\bsub{Real Time Clock (RTC).}
RTC provides timers by maintaining an accurate time base via the crystal oscillator.
As clocks usually have fixed drifts from ideal frequencies, we use this feature (i.e., \textbf{RTCFre}) to set timers with diverse frequencies to statistically record the accumulated time drift. 
The time phase~\cite{time-stamp} among multiple clocks can also be used as a feature (i.e., \textbf{RTCPha}).
On MCU-based devices, the main clock is always a fast clock and peripheral clocks are always slow. 
For instance, the main clock's frequency is 180MHz and a crystal oscillator clock's is 32kHz on STM32F429. 
Thus, we can use the dual clocks of the system and the peripherals to get instantaneous phases and measure the phase features. 

\bsub{SRAM.}
Previous approaches~\cite{sram-puf-white-paper, puf-taxonomy} indicate that the initial states of SRAM cells are usually stable and can be used as a kind of PUF.
We collect the initial bit states within a specified SRAM address range during device boot-up and employ statistical features derived from these bits to differentiate various devices.

\begin{table}[t]\color{mytablecolor}
\centering
\caption{\modify{Hardware modules on STM32F4 serials with their functional descriptions and sources of uniqueness.}}
\label{tab:hardware_modules}
\begin{tabular}{@{}ccc@{}}
\toprule
\textbf{Hardware on MCUs*} & \textbf{Functionality} & \textbf{Source of Uniqueness} \\ \midrule
TIMx. RTC, WDG       &    Clock and timer        & Skew\cite{IoT-ID}, Phase\cite{time-stamp}    \\ \midrule
SRAM, Flash, ROM &  Storage medium & Special property\cite{flash-memory} 
\\ \midrule
ADC, DAC, PWM & Voltage processor & Numerical error\cite{IoT-ID}                                \\ \midrule
FPU, CRC, CRYP & Computing units   & Performance\cite{drawnapart_gpu}                           \\ \midrule
PWR, DMA, RCC   &  System controller      & Manufacturing defects\cite{BLE/Tracking}
\\ \midrule
I2C, SPI, USART & Data transmitter       & Transmission delay\cite{ECU} 
\\ \bottomrule
\multicolumn{3}{l}{* All abbreviations refer to \textit{RM0090 Reference Manual}.}   
\end{tabular}
\end{table}

\bsub{Flash.}
Previous study~\cite{time-print} uses NAND-flash's different sector read times as distinctive fingerprinting features because the NAND-Flash is located on a separate chip with dedicated drivers, which requires access via the I2C/SPI bus and can affect the access time of sectors. 
In contrast, most MCUs are only equipped with NOR-Flash, where different sectors exhibit similar read times due to their integration on the same chip as the MCU, enabling direct access.
Thus, we exclude this feature as only a few devices have NAND-flash.



\begin{figure}[t]
    \centering
    \subfloat[Mapping in DAC-ADC\label{fig:design_sample_mapping}]{\includegraphics[width=0.24\textwidth]{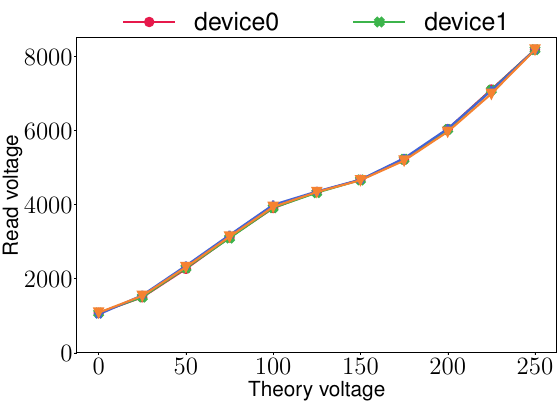}}
    \hfill 	
    \subfloat[Density of DAC-ADC\label{fig:design_sample_distribution}]{\includegraphics[width=0.24\textwidth]{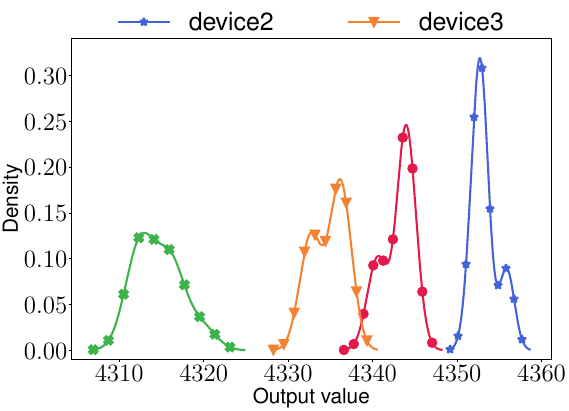}}
    \caption{Example of DAC-ADC fingerprints of four ESP32S2 devices. (a) shows the mapping between theory voltages of DAC and read voltages of ADC. (b) shows the density distributions of the read voltages whose theory voltage is 125.}
    \label{fig:design_sample}
\end{figure}

Rather than generating a static fingerprint, we manipulate the settings or inputs to generate varying fingerprints from these hardware features. 
Each individual hardware feature can serve as a fingerprinting task, producing multiple fingerprint results by utilizing different input arguments.
For instance, we use DAC to generate a voltage and ADC to read it. 
Theoretically, the read voltage of ADC and the input voltage of DAC can be formulated as a proportional mapping,
\begin{equation}
\label{eq:DAC_ADC_sample}
V_{ADC} = V_{DAC} * \frac{2^{res_{ADC}} - 1}{2^{res_{DAC}} - 1}
\end{equation}
$res$ means the resolution. 
However, as shown in Figure~\ref{fig:design_sample_mapping}, this mapping is not exactly linear and its density distribution may vary by devices (see Figure~\ref{fig:design_sample_distribution}).
By inputting different voltages via DAC, we can distinguish different devices from the actual voltage read by ADC.
Similar approaches can be applied to other hardware features, such as setting different RTC clock sources and reading different address ranges in SRAM. The detailed designs of the fingerprinting tasks for the 6 hardware features are discussed in Appendix~\ref{chp:hardware_fingerprints_details}.
In this way, we can gain enough $(arguments, fingerprint)$ pairs for different requests.







\subsection{Binding Requests with Unique Hardware Fingerprints}
\label{sec:message_mapping}

We aim to protect the integrity of client requests, so as to prevent adversaries from gaining access to the backend via tampering with users' requests.
Traditional techniques based on message authentication codes~\cite{iot_auth, iot_ssl} are not sufficient, since they require the client side to store secret keys which would be easily compromised on IoT devices as reported by~\cite{iot_hazards, share_keys}.
We propose to bind each request with unique hardware fingerprints without relying on the assumptions of key security on IoT devices. In this way, any deliberate manipulation targeting requests would be detected by checking the correctness of the fingerprints.







\begin{algorithm}[t]
    \caption{Message Mapping Algorithm}
    \label{alg:mapping}
    \KwIn {$request$}
    \SetAlgoLined
    \KwOut {$task$}
    $task \leftarrow [\ ],$ $digest \leftarrow 0$\\
    $operation \gets$ GetOperation($request$) \\ 
    $nonce \gets$ GetNonce($request$) \\
    $payload \gets$ GetPayload($request$) \\
    $totalNum \gets$ GetTaskNumber() \\ 
    \For {$i \gets range(0, totalNum)$} {
        \text{\small{// provide protection for the operation}} \\
        $h_1 \gets$ Hash($operation,nonce, digest$)\\
        \text{\small{// provide protection for the payloads}} \\
        $h_2 \gets$ Hash($nonce, payload_i$)\\
        \text{\small{// connect with another payload}} \\
        $h_3 \gets$ Hash($nonce, payload_{\textnormal{-}i}$) \\
        $digest \gets$ Hash($h_1,h_2,h_3$)\\
        $arg_0, ..., arg_m \gets$ DivideArguments($digest$)\\
        $task_i \gets (arg_0, ..., arg_m)$
    }
    \Return $task$ \\
\end{algorithm}

As described in \S~\ref{sec:choose_feature}, we leverage multiple hardware modules to execute some hardware tasks which take as input the request-derived arguments and output unique hardware fingerprints.
Here, the request should be mapped into unique task arguments to avoid collisions between fingerprints from different requests.
Unfortunately, the requirement may be hard to satisfy, because the input space of one hardware task may be very small.
{As an example, the number of the arguments designed for ESP32S2 is about 20,000.
For a hash function whose outputs are mapped into a space size (denoted as $d$) of 20,000, assuming there are $2^{10}$ distinct inputs, the probability of a collision in the output has already reached 99\%.
}
%
If each task's arguments are generated by the above hash function based on the specific content from the request, an attacker can compromise the integrity of the request with high probability by manipulating specific content and constructing collisions.




To address the issue, we devise a novel random message mapping algorithm that maps some content in the request to arguments of multiple hardware tasks via a hash function, rather than each content corresponding to only one task.
This exponentially increases the output space of the single hash function, thus exponentially reduces the probability of collisions.
%
%
%
%
The detailed process is shown in Algorithm~\ref{alg:mapping}.
$request$ contains an operation, a nonce (e.g., a random number), and several payloads (line 2-4). 
For each $request$, we divide payloads into \totalnumber groups and generate the $i$-th payload via $payload[i\ mod\ totalNum]$ sequentially.
Note that \totalnumber means the number of tasks used for an authentication which is a pre-defined fixed number (line 5). 
Then, we generate task arguments for the \totalnumber hardware tasks according to the $request$ information (line 6-15). For each round, $h_1$ is the digest of the operation, the nonce, and the digest in the last round. 
$h_2$ is the digest of the $i$-th payload from the beginning of the payload group  while $h_3$ is calculated by the $i$-th payload from the end of the payload group.
Such a design correlates a payload content to arguments of two hardware tasks, decreasing the probability of output collisions.
Finally, we concatenate $h1, h2, h3$ and use an extra hash calculation to get the $digest$ in this round.
The specific segmentation of $digest$ constitutes the arguments which are further fed into the corresponding hardware tasks.

\subsection{Countering Machine Learning Attacks}
\label{sec:data_poisoning}

Once requests and fingerprints are transferred in the network, they may be abused by attackers.
Essentially, they may learn the relationships between requests and fingerprints and then forge reasonable fingerprints to cheat the backend (i.e., software mimic attacks).


To resist those misbehaviors, one intuitive method is to make the hardware tasks as complex as possible to prevent the attackers from easily learning the relationships.
Existing works on PUF~\cite{puf_suervey} concentrate on constructing  unpredictable relationships between the inputs and outputs of hardware modules.
However, these approaches may depend on special circuits which are only available on some dedicated devices.
In this work, instead of relying on specific complex hardware tasks, we tend to generate unlearnable fingerprints only based on some common hardware modules (as specified in \S~\ref{sec:choose_feature}).

Inspired by the data poisoning attacks widely explored in the area of machine learning~\cite{data_poison}, we randomly add some well-crafted noises to the raw hardware fingerprints to generate poisoned fingerprints.
There are three requirements for the poisoned fingerprints:
(1) Verifiability: the poisoned fingerprints can be successfully authenticated by the backend according to the raw ones.
(2) Dissimilarity: the poisoned fingerprints should detach from the raw ones as much as possible to prevent attackers from learning the features of the raw fingerprints.
(3) Unidentifiability: the noises in those fingerprints cannot be identified and removed by attackers through advanced techniques such as machine learning.


To satisfy the first requirement, we randomly retain a portion of the raw fingerprints as normal ones which will be used to pass the authentication on the backend side.
For the remaining fingerprints, we make a trade-off between the dissimilarity and the unidentifiability when adding random noise.
To increase the dissimilarity between the raw and the poisoned fingerprints, the noises added should be as large as possible yet would enable attackers to easily identify poisoned fingerprints.
Here, we generate the well-crafted noises that are slightly larger than the inherent hardware errors.
%
%
Specifically, for a pair of $(arguments, fingerprint)$, the poisoned fingerprint is computed as:
\begin{equation}
\label{eq:poison_strategy}
fp_{poisoned} = fp_{raw} * (noise + 1) + C
\end{equation}
where $C$ is a constant and $noise$ is randomly sampled from distributions (e.g., Laplace distribution).

\subsection{Verifying Fingerprints at Backend}
\label{sec:backend_verifier}
We describe how to authenticate a request along with its fingerprints on the backend side.
According to the above construction method of poisoned fingerprints, a straightforward solution  is to compare them with the raw fingerprints and check whether the number of matched elements is larger than a pre-defined threshold. 
\todo{B1}
\modify {
Based on this idea, we propose a novel fingerprint predictor to mimic the behavior of each hardware module and to predict an approximation of the raw fingerprints generated by the clients.    
The predicted fingerprints generated by the predictor are finally fed into a verifier to compare with the poisoned fingerprints sent by the clients.
Throughout the authentication process, the backend does not know if a fingerprint is poisoned.
It just checks the number of fingerprints that match the raw ones to decide whether the authentication passes.
The retained raw fingerprints (i.e., the normal ones) will match the predicted ones, ensuring that the authentication can succeed.
The details are as follows.
}



The predictor consists of a set of sub-predictors, which are regression models for one task of one client.
Essentially, before one client device is deployed \modify{(e.g. during device manufacturing in a factory)}, the backend collects enough $(arguments, fingerprint)$ pairs for each task and uses them to train a new sub-predictor. 
Meanwhile, the verifier also contains a set of sub-verifiers, each as a  binary classifier for one sub-predictor.
In the deployment phase, a sub-verifier is trained using fingerprints from the corresponding client (same with the sub-predictor's) as positive samples and those from other clients as negative samples.
In the real-world environment, it takes as input one predicted fingerprint generated by the corresponding sub-predictor and that from a client and outputs whether the latter is correct.

When receiving a request from one client, the authentication process has the following steps.
(1) The backend uses the message mapping algorithm to generate \totalnumber tasks, similar to the relevant operations on the client side.
The task output item is represented by a $(arguments, fingerprint)$ pair for convenient utilization in later operations.
(2) The predictor launches appropriate sub-predictors for the above tasks and predicts the corresponding fingerprints based on the tasks and arguments.
(3) The verifier uses relevant sub-verifiers to check whether the predicted fingerprints match that of the client for each task.
The backend determines authentication as valid by checking if the number of matched fingerprints exceeds a pre-defined threshold (i.e., \acceptnumber).
Specifically, the backend maintains a timestamp or sequence number of the requests to prevent replay attacks.
To prevent wireless signal relay, the backend measures the message round trip time and compares it with the predicted times based on specific tasks.
(4) The backend returns the authentication result to the client for further communications.

\section{Security Analysis}
\label{chp:sec_analysis}




\subsection{Countering Fingerprint Mimic Attacks}


In this subsection, we present how \ourwork~resists fingerprint mimic attacks, including hardware mimic attacks and software mimic attacks as introduced in \S~\ref{sec:threat_model}.


\subsubsection{\textbf{Hardware Mimic Attack}}
To launch hardware mimic attacks, one adversary may purchase devices with \ourwork~installed and with the same types of hardware as the victim devices to generate fingerprints for any request.
Regarding those attacks, we utilize hardware features to construct hardware fingerprints. These hardware features exhibit variations across different devices, even if they have the same model, making fingerprints generated for different devices distinct. Consequently, attackers' devices can be identified as unauthorized due to the distinctive fingerprint patterns derived from the specific hardware characteristics.




\subsubsection{\textbf{Software Mimic Attack}}
Attackers can launch software mimic attacks by eavesdropping on communication channels, monitoring requests with fingerprints, and learning their relationships.
To defeat those misbehaviors, \ourwork~utilizes the data poisoning based method that adds random well-crafted noises to the raw hardware fingerprints. 
Note that the noises not only make adversaries fail to learn the correct relationships between the requests and the raw hardware fingerprints, but are also random and stealthy to avoid adversaries identifying and removing them.
It is difficult for attackers to distinguish the poisoned data as the data is slightly modified from valid fingerprints.
This discrepancy is adequate to prevent attackers from precisely predicting fingerprints.
If attackers learn with poisoned fingerprints, the deviated data can significantly degrade the performance of their model.
We present a quantitative analysis through a linear hardware task as an example, which can be represented by the function:
When all the fingerprints are poisoned, the function parameters fitted by an attacker are,
\begin{equation}
\label{eq:poison_result}
\begin{aligned}
w' & = (1+noise)*w\\
b' & = (1+noise)*b  + C
\end{aligned}
\end{equation}
\modify{
In Appendix~\ref{chp:proof_poison}, we prove the above equation and show how \ourwork rejects the poisoned fingerprints.
}




\subsection{Countering Fingerprint Reuse Attacks}

Besides fingerprint mimic attacks, we illustrate the security of \ourwork~against fingerprint reuse attacks, including replay attacks, forwarding attacks, tampering attacks, and relay attacks.


\subsubsection{\textbf{Replay Attack}}
In our message mapping approaches, any changes in the request payload result in different fingerprints. 
Therefore, \ourwork can utilize the existing timestamp or sequence numbers of the protocols, or keep our nonce growing.
The backend can record the last value of this increasing number and reject repeated requests.


\subsubsection{\textbf{Relay Attack}}
For wireless devices, attackers may relay the physical signals (e.g., BLE~\cite{ble_relay_attack}, RFID~\cite{pke_relay}) to valid devices at a distance.
Similar to existing approaches~\cite{reid2007detecting, singh2017uwb}, \ourwork can measure the request's round-trip time to identify the signal relay.
For the requests delivered by networks (e.g., HTTP), attackers may forward authentication requests to valid devices. 
Our one-round protocol can ensure that requests are initiated by the clients, so attackers cannot simply offload the server's requests to trigger authentication on real devices.
They can only try to reuse the fingerprints of existing requests, which is discussed in the subsequent tampering attacks.


\subsubsection{\textbf{Tampering Attack}}
If an attacker tampers with requests to launch tampering attacks, such as replacing operations in the request while retaining the fingerprints, \ourwork~can easily detect such misbehaviors with high probability. 
Specifically, the request is tightly bound with its fingerprints via the message mapping algorithm.
Any unwanted modifications targeting request contents would result in significantly different hardware tasks being generated on the server and client sides using the same message mapping algorithm with a high probability.
Compared to the naive approach of just hashing once, our algorithm can exponentially increase the attackers' attempt times.
\modify{
In Appendix~\ref{chp:proof_tamper}, based on a hash collision problem, we analyze Algorithm~\ref{alg:mapping} step by step to show how \ourwork creates obstacles to the tampering attack.
}

\section{Evaluation}

To evaluate \ourwork's authentication performance and security, we aim to answer the following four questions:


\begin{enumerate}[label=\textbf{Q\arabic*}]
\item Which hardware features can be used for fingerprinting?
\item How accurate is \ourwork's~authentication under different client and backend settings? 
\item Can \ourwork defend against various fingerprint mimic attacks and reuse attacks? 
\item How much overhead \ourwork bring to real scenarios? 
\end{enumerate}

To answer \textbf{Q1}, we evaluate the performance of every single fingerprint and their combinations for device authentication and their stability under different environments (\S~\ref{chp:eval_auth}).
For \textbf{Q2}, we show the true positives and false positives of \ourwork~in authentication  with different parameter settings, especially poisoning-related configurations
(\S~\ref{chp:eval_ourwork}).
For \textbf{Q3}, we launch hardware mimic attacks, software mimic attacks, and tampering attacks to evaluate the security of  \ourwork (\S~\ref{chp:eva_attack}).
At last, we conduct case studies on various usage scenarios to demonstrate the usability of \ourwork~for \textbf{Q4} (\S~\ref{chp:eva_overhead}). 


\subsection{Experiment Setup}

\subsubsection{\textbf{Client-side Implementation}}
We implement \ourwork on $30$ ESP32S2, $20$ STM32F103, and $10$ STM32F429 MCU-based devices, as shown in Table~\ref{tab:devices}.
We deploy the 6 hardware features in \ref{sec:choose_feature} on these devices.
Details of the designed tasks are shown in Appendix~\ref{chp:hardware_fingerprints_details}.

\begin{table}[t]
    \caption{Devices used in evaluation.}
    \label{tab:devices}
    \centering
    \begin{tabular}{@{}cccc@{}}
    \toprule
    \textbf{Model-brand} & \textbf{Microcontroller} & \textbf{Frequency} & \textbf{\# of devices} \\ \midrule
    ESP32S2     & Xtensa LX7      & 240MHz    & 30     \\ \midrule
    STM32F103   & Cortex M4       & 72MHz     & 20     \\ \midrule
    STM32F429   & Cortex M4       & 180MHz    & 10     \\ \bottomrule
    \end{tabular}
\end{table}


\subsubsection{\textbf{Backend-side Implementation}}
We implement the backend authentication service using Python and deploy it on a Windows 10 PC with 16 GB RAM and 2.8 GHz CPU.
The communication between the backend and the client devices are developed through serial ports.
Our predictors are regression models, specifically ExtraTrees, and our verifiers are classification models, specifically RandomForest, all implemented using Scikit-learn \cite{scikit-learn}.
\todo{A3}
\modify{
The hash function used in Algorithm~\ref{alg:mapping} is APHash\footnote{\href{https://github.com/ArashPartow/hash}{https://github.com/ArashPartow/hash}}{https://github.com/ArashPartow/hash}.
For data poisoning, $noise$ is obtained from the uniform distribution of [0.08,0.2], and we empirically set $C$ (in Equation~\ref{eq:poison_strategy}) to 1 (discussed later). 
}


The regression models are trained with $(arguments,$ $fingerprint)$ pairs in the model training phase.
When training classification models, we randomly sample 10 other devices as negative examples.
We train a regression model and a classification model for each hardware task of each client.
{For each device and hardware feature, we gather 5,000 pairs of data. We use half of them to train our models and the other half to test.}



\subsection{Usability of Hardware Feature for Fingerprinting}
\label{chp:eval_auth}
\subsubsection{\textbf{{Identifiability of Hardware Fingerprints for Device Authentication}}} 

We evaluate whether a hardware feature can be used to identify one device in the \modify{subsection}.
For each hardware task on each device, we generate a fingerprint (without poisoning) and check two properties:
(1) if it can be successfully verified at the backend side;
(2) {if it will be misidentified as other devices}.
In the evaluation, we use all the devices described above. For each device type, we separately employ each device to impersonate all other devices to figure out whether it will be misidentified as other devices.
Especially, we build two metrics: (1) true positive rate (TPR) {equal to the proportion of devices that pass the authentication process successfully to the total devices}, (2) false positive rate (FPR) {equal to the proportion of misidentified devices to the total devices}.


Table~\ref{tab:eva_closed} shows the TPRs and FPRs of different types of features of different devices.
We can see that RTCFre and SRAM achieve a TPR of more than \modify{90\%} and an FPR of less than \modify{8\%} for various devices, meaning that these two hardware features can identify one device with a high probability.
By contrast, there are hardware features with high FPRs, such as the FPU on ESP32 and STM32F103.
\modify{
\todo{C5}
FPUs are unavailable on ESP32S2 and STM32F103.
We use software-based floating point calculators, which results in a high false positive.
}

Besides evaluating each individual hardware features, we {utilize multiple hardware features to achieve more accurate authentication. We eliminate useless fingerprints for each device category, i.e., FPU and RTCPhra for ESP32S2, PWM and RTCPhra for STM32F249, DAC/ADC, FPU and PWM for STM32F103. 
As shown in Table~\ref{tab:eva_closed}, our approach achieves a FPR of \modify{1.06\%} while maintaining a TPR of \modify{98\%} (on ESP32S2).} 


\begin{table}[t]
\centering
\caption{TPRs and FPRs of various hardware features.}
\label{tab:eva_closed}
\begin{threeparttable}
\begin{tabular}{@{}ccccccccc@{}}
\toprule
          & \multicolumn{2}{c}{\textbf{ESP32S2}} &  & \multicolumn{2}{c}{\textbf{STM32F429}} &  & \multicolumn{2}{c}{\textbf{STM32F103}} \\ \midrule
          & TPR          & FPR          &  & TPR           & FPR           &  & TPR           & FPR           \\ \midrule
DAC\_ADC         & \red{83.74}       & \red{8.58}         &  & \red{82.73}         & \red{16.83}         &  & \red{96.25}         & \red{37.90}         \\
FPU             & \red{76.59}        & \red{38.90}         &  & \red{83.50}         & \red{29.94}         &  & \red{76.65}         & \red{36.63}         \\
PWM             & \red{84.83}        & \red{17.54}         &  & \red{84.90}         & \red{37.67}         &  & \red{80.00}         & \red{35.57}         \\
RTCFre          & \red{91.76}        & \red{1.96}         &  & \red{89.88}         & \red{7.49}         &  & \red{99.19}         & \red{1.96}         \\
RTCPha          & \red{77.04}        & \red{58.38}         &  & \red{73.88}         & \red{58.10}         &  & \red{74.56}         & \red{36.88}         \\
SRAM            & \red{94.27}        & \red{0.01}         &  & \red{98.69}         & \red{0.05}         &  & \red{96.89}         & \red{0.03}         \\ \midrule
Ensemble  & \red{96.63}        & \red{9.44}        &  & \red{97.06}         & \red{14.10}         &  & \red{97.94}         & \red{14.31}         \\
Ensemble* & \red{98.47}        & \red{1.06}         &  & \red{97.67}         & \red{6.89}          &  & \red{98.68}         & \red{1.64}          \\ \bottomrule
\end{tabular}
\begin{tablenotes}
\item[*] The results of excluding useless features, \modify{i.e., FPU and RTCPhra for ESP32S2, PWM
and RTCPhra for STM32F249, DAC/ADC, FPU and PWM for STM32F103.}
\end{tablenotes}
\end{threeparttable}
\end{table}

\subsubsection{\textbf{Stability of Hardware Fingerprints under Various Real-world Environments}}
\begin{figure}[t]
    \centering
    \includegraphics[width=0.40\textwidth]{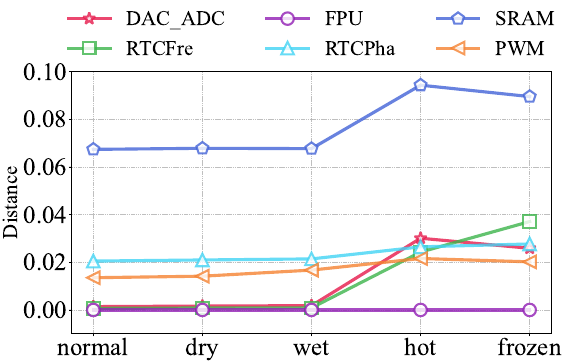}
    \caption{Features fluctuations in different environments.}
    \label{fig:env}
\end{figure}

We evaluate the stability of hardware features in different environments with varied temperatures and humidity. 
The environmental parameters in normal conditions are 28$^{\circ}$C and 61\% relative humidity (RH).
Besides, we set up two humidity environments: 37\% RH (called $dry$) and 98\% RH (called $wet$), and two temperature environments: $-23$$^{\circ}$C (called $frozen$) and 52$^{\circ}$C (called $hot$).
{We collect $(arguments, fingerprint)$ pairs under different environmental conditions. Then, we calculate the distances between these pairs and the pairs collected under normal conditions. The distances are calculated as the relative error between $fingerprint$s under the same $arguments$. }
%
The average distance fluctuations during different environments are shown in Figure~\ref{fig:env}.

We find that fluctuations of different hardware features are different.
For all features, the degree of change in distances is less than 0.1. 
Considering the influence of the two factors, humidity has almost no influence.
It is evident that temperature has a significant impact on hardware fingerprints, in particular on RTCFre and DAC/ADC. However, it is important to note that these temperature settings are rarely encountered in real-life scenarios. Even if they do exist, we can collect fingerprint information from these environments to train the backend's predictors and validators to avoid false positives/negatives.



\subsection{Authentication Accuracy of \ourwork}
\label{chp:eval_ourwork}
We assess the authentication accuracy of \ourwork under different parameter settings.
There are three parameters in the work: 
(1) \totalnumber denotes the number of hardware tasks executed by  clients; 
(2) \usednumber represents the number of fingerprints without being poisoned by clients; 
(3) \acceptnumber depicts the threshold of verified fingerprints required for successful device authentication at the backend side.
In the evaluation, we select {DAC/ADC, PWM, RTCFre, and SRAM} 
in \ourwork, according to the evaluation results in \S~\ref{chp:eval_auth}.
By default, we set \totalnumber to 10, \usednumber to 5, \acceptnumber to 3.
The following evaluations are all based on ESP32S2 devices.



\begin{figure}[t]
    \centering
    \subfloat[Different \usednumber\label{fig:auth_used_number}]{\includegraphics[width=0.23\textwidth]{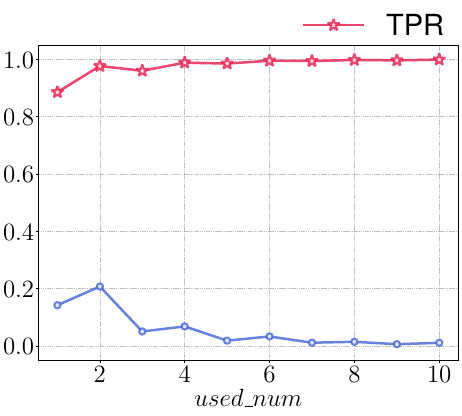}
    }
    \hfill 	
    \subfloat[Different \acceptnumber\label{fig:auth_accept_number}]{\includegraphics[width=0.23\textwidth]{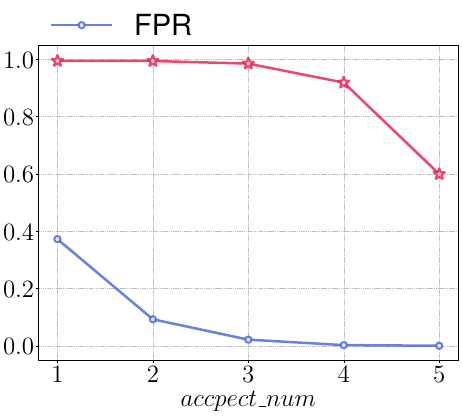}}
    \caption{TPRs and FPRs of \ourwork under different parameter settings.}
    \label{fig:auth_ourwork}
\end{figure}

We set $totalNum=10$ and change other two parameters.
Figure~\ref{fig:auth_used_number} shows the results under varied \usednumber and $acceptNum=\lceil \frac{usedNum}{2} \rceil$.
{When $usedNum$ is 1, \ourwork only uses one type of hardware feature as the fingerprint, the TPR or FPR  is equal to the average TPR or FPR value of all individual features in Table~\ref{tab:eva_closed}.
As \usednumber increases, TPR increases and FPR decreases because the larger number of used fingerprints provides more information of device identities.
In Figure~\ref{fig:auth_accept_number}, we change \acceptnumber with \usednumber as $5$.
As \acceptnumber increases, more fingerprints need to be verified by the verifier, leading to the reduction in both TPR and FPR.
}
%
Actually, we can set the ratio of \acceptnumber to \usednumber to balance the TPRs and FPRs. 



\begin{figure*}[t]
    \centering
    \begin{minipage}[c]{0.25\textwidth}
        \includegraphics[width=0.9\textwidth]
        {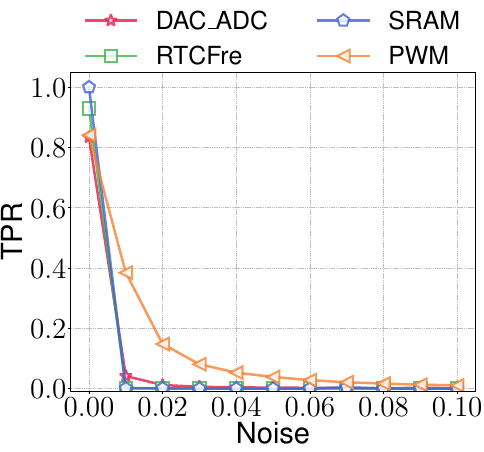}
        \caption{Authenticate with poisoned fingerprints. TPR shows whether authentications pass.}
        \label{fig:attack_single_noise}
    \end{minipage}
    \hfill
    \begin{minipage}[c]{0.7\textwidth}
    \centering
    \subfloat[Different \usednumber\label{fig:attack_multiple_used_number}]{\includegraphics[width=0.32\textwidth]{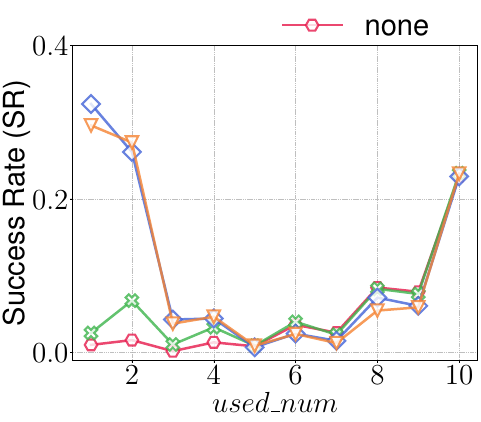}}
    \hfill 	
    \subfloat[Different \acceptnumber\label{fig:attack_multiple_accept_number}]{\includegraphics[width=0.32\textwidth]{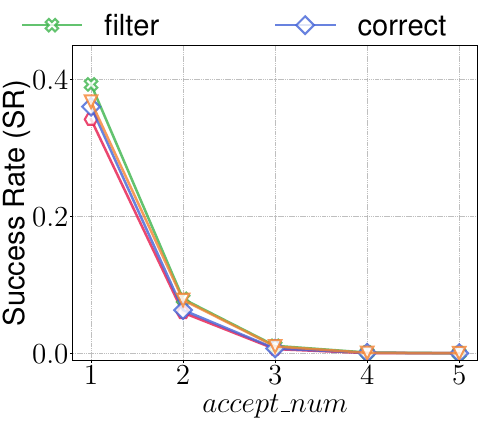}}
    \hfill 	
    \subfloat[Different used ratio\label{fig:attack_multiple_rate}]{\includegraphics[width=0.32\textwidth]{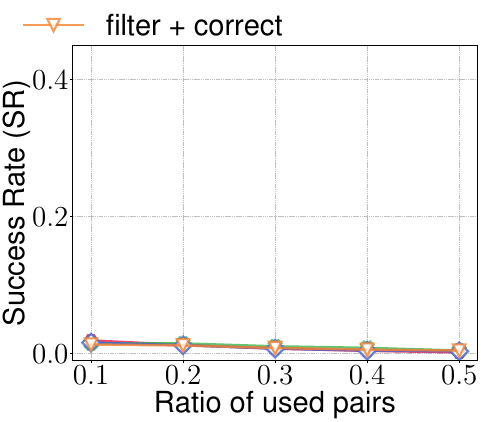}}
    \hfill 	
    \caption{Software mimic attack results on \ourwork. "filter" indicates that the attacker filters out poisoned pairs for training and "correct" indicates the attacker corrects outputs.}
    \label{fig:attack_ourwork}
    \end{minipage}
\end{figure*}

\modify{
\todo{A2}
To confirm whether the normal fingerprints can pass authentication and the poisoned ones cannot, we conduct experiments by introducing various levels of $noise$ to modify the raw fingerprints.
As shown in Figure~\ref{fig:attack_single_noise}, low TPRs indicate that the poisoned fingerprints are unlikely to pass authentication.
When $noise$ exceeds 0.08 (the $noise$ used is from [0.08, 0.2]), the TPR drops to less than 2\%.
The results show that successful authentication only relies on normal fingerprints and is not affected by poisoned ones.
}

\begin{table}[t]
\centering
\caption{Hardware mimic attack success rates.}
\label{tab:attack_hardware}
\begin{tabular}{@{}cccc@{}}
\toprule
          & {ESP32S2} & {STM32F103} & {STM32F429} \\ \midrule
ESP32S2   & 0.0188  & 0.0000    & 0.0000    \\
STM32F103 & 0.0001  & 0.0606    & 0.0078    \\
STM32F429 & 0.0000  & 0.0000    & 0.1058    \\ \bottomrule
\end{tabular}
\end{table}

\subsection{Security of \ourwork against Various Attacks}
\label{chp:eva_attack}
We launch various adaptive attacks including hardware mimic attacks, software mimic attacks, and tampering attacks to evaluate the security of \ourwork.

\subsubsection{\textbf{Hardware Mimic Attack}} 
During the experiment, we simulate hardware mimic attacks by having an attacker use a device that has the same or similar brand and model as the victim's device. 
The devices are randomly divided into two groups: legitimate devices and attacking devices. 
We then initiate the authentication process using the attacking devices and assess whether they are correctly identified as illegal devices. 
Finally, we measure the success rate of impersonation attacks using various types of devices.
The success rate is shown in Table~\ref{tab:attack_hardware}.

In Table~\ref{tab:attack_hardware}, the rows represent the type of devices that are known to the backend (called target devices) while the columns represent the types of devices used by the attacker (called source devices).
When the source device has the same brand and model as the target device, the attacker can successfully launch an attack.
However, the success rates are still low, less than 11\%.
For the attacks using different device models, the success rates are even lower, with less than 0.01\% success rate.
These results indicate that hardware mimic attacks are ineffective against \ourwork.

\subsubsection{\textbf{Software Mimic Attack}} 
We evaluate the machine learning based software mimic attacks and consider the that attackers can collect $(arguments, fingerprint)$ pairs (which are partially poisoned) and train a regression model to learn the relationship between $arguments$ and $fingerprint$.
First, we evaluate the effectiveness of \ourwork in defending against software mimic attacks.
Then, we show the effectiveness of \ourwork by analyzing attacks on single features.
Furthermore, we consider an attacker who attempts to filter out poisoned pairs to demonstrate that poisoned pairs are unable to be identified.



\noindent \textbf{Defending Effectiveness Against Software Mimic Attacks}.
We evaluate the effectiveness of \ourwork in defending against machine learning attacks with different attack settings. 
The backend authentication service's settings are the same as \S~\ref{chp:eval_ourwork}. 
We consider an attacker who trains a regression model for each hardware feature and generates fake fingerprints based on the requests to cheat the backend authentication.
\modify{The models used by attackers are the same as those used by the backend.}
Furthermore, attackers can employ various training and \modify{predicting} strategies to carry out their attacks.
For training, the attacker chooses to filter $(arguments,fingerprint)$ pairs,
(1) the attacker uses all the pairs directly;
(2) the attacker randomly selects \usednumber pairs as normal pairs.
At the same time, we consider that the attacker may try to correct the output fingerprints,
(1) the attacker predicts fingerprints directly;
(2) the attacker predicts fingerprints directly with probability $usedNum/totalNum$. Otherwise, the attacker selects a value from the $noise$ range and corrects fingerprints through the reverse process of poisoning. The attacks with corrected results can utilize the poisoned pairs to improve attack performances.
Combining the choices of filtering and correcting, there are 4 different attack strategies.
We evaluate success rates for different attack strategies under various parameter settings.
The results are shown in Figure~\ref{fig:attack_ourwork}.

\begin{figure*}[t]
    \centering
    \subfloat[Authenticating without protection\label{fig:attack_single_without_protect}]{\includegraphics[width=0.24\textwidth]{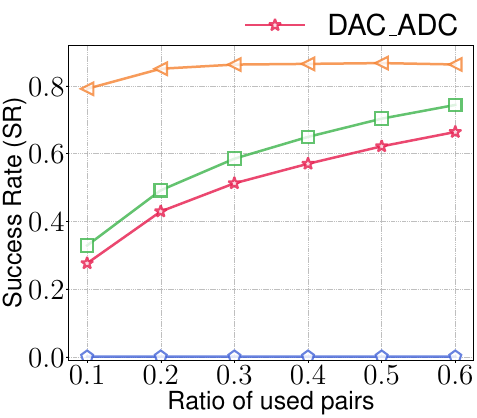}}
    \hfill 	
    \subfloat[Authenticating with protection\label{fig:attack_single_with_protect}]{\includegraphics[width=0.24\textwidth]{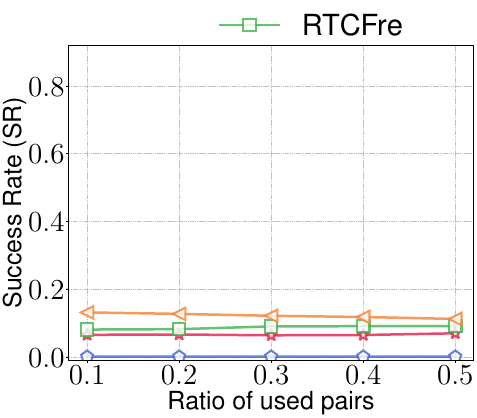}}
    \hfill 	
    \subfloat[Different \usednumber\label{fig:attack_single_used_num}]{\includegraphics[width=0.24\textwidth]{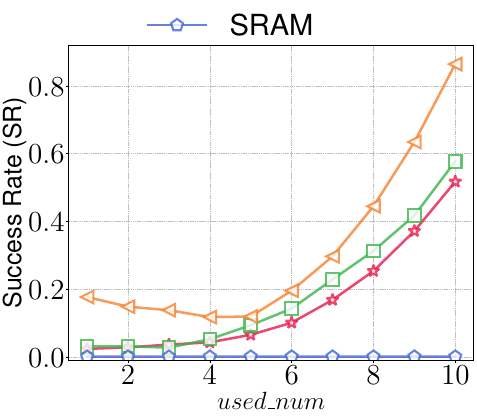}}
    \hfill 	
    \subfloat[Different \acceptnumber\label{fig:attack_single_accept_num}]{\includegraphics[width=0.24\textwidth]{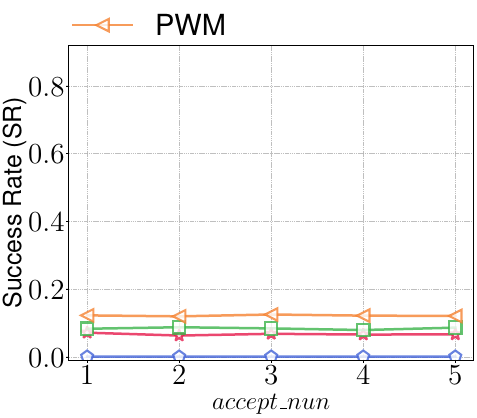}}  
    \caption{Software mimic attack success rates on single fingerprints. (a) and (b) are the attack performances without and with data poisoning, respectively. (c) and (d) demonstrate the influence of different parameter settings.}
    \label{fig:attack_single}
\end{figure*}

The parameter settings are the same as those used for device verification. The used ratio of every single feature is 30\%, which means during the attacks the attacker obtains 30\% of all normal pairs (non-poisoned pairs) for training.
The value of \usednumber determines the ratio of the normal pairs.
In Figure~\ref{fig:attack_multiple_used_number} we vary \usednumber from 1 to 10.
When \usednumber is 1 or 2, the majority of pairs obtained by the attacker are poisoned and the authentication success threshold (i.e., \acceptnumber) is set to a very low value (\modify{i.e.}, $1$).
Therefore, the strategies involving corrected output achieve a high success rate of 32.3\%.
When \usednumber is close to \totalnumber, the attacker obtains more normal pairs and trains his models more effectively.
When the number of normal pairs and poisoned pairs is equal, the entropy is at its highest, resulting in an attack success rate of around 1\%  regardless of the strategies.
Figure~\ref{fig:attack_multiple_accept_number} shows that as \acceptnumber grows, the difficulty of passing the authentication also increases.

In Figure~\ref{fig:attack_multiple_rate}, we change the used ratio of normal pairs.
The used ratio refers to the number of normal pairs obtained by the attacker.
We find that when the attacker gets more normal pairs, the success rate (SR) decreases but does not increase.
\modify{The reason} is that with more normal pairs there are more poisoned pairs (with the default setting, the number of normal pairs and poisoned pairs are the same).
The models trained by the attacker are affected more, resulting in the generation of invalid fingerprints for the corresponding arguments.
This indicates that data poisoning is effective in preventing software mimic attacks.



\noindent \textbf{Software Mimic Attacks on Single Features.} 
We further analyze the mimic attacks on single features.
In this experiment, we use the same attack settings as in the previous experiment. 
\modify{
The training process for the attackers is the same as before.
For testing, we use only one fingerprint for authentication and check if the backend is fooled.
We use the highest success rate of the four attack strategies as the final result.
}


Figure~\ref{fig:attack_single_without_protect} and Figure~\ref{fig:attack_single_with_protect} show how \ourwork provides protection to a single feature.
In Figure~\ref{fig:attack_single_without_protect}, we set $usedNum=totalNum$, and the pairs obtained by the attacker are all normal ones.
In Figure~\ref{fig:attack_single_with_protect}, we set $usedNum:totalNum=1:2$, and half of the pairs are poisoned.
It is important to note that, in these two different settings, the number of obtained normal pairs is the same, but in the latter one there are extra poisoned pairs.
Without protection, the success rate mainly depends on complexity of the features.
For SRAM, the power-on voltages of SRAM cells are unpredictable so the success rate is very low (almost 0\%) no matter how many pairs are known.
But for other single features, the attacker achieves more than 50\% success rate with 0.3 of all the normal pairs, particularly in the feature PWM. 
When protected by \ourwork, the success rate on PWM decreases to approximately 13\% and for other fingerprints, the success rate is lower than 10\%.
The magnitude of the decline is remarkable.
More importantly, as the obtained ratio increases, the success rate decreases.
The results prove that the presence of poisoned pairs helps protect single features.

As for the parameters of \ourwork, \usednumber affects the ratio of normal pairs.
The attack success rate for a single feature will initially decrease and then increase as \usednumber increases.
\acceptnumber only works when authenticating with multiple features and has no influence on a single feature.
The results are shown in Figure~\ref{fig:attack_single_used_num} and Figure~\ref{fig:attack_single_accept_num}.

\begin{table}[t]
\caption{Identification accuracy of poisoned fingerprints.}
\label{tab:poison_data_identify}
\centering
\begin{tabular}{@{}ccccc@{}}
\toprule
             & \textbf{DAC/ADC} & \textbf{RTCFre} & \textbf{SRAM}   & \textbf{PWM}    \\ \midrule
Unsupervised learning      & 0.5201  & 0.5042 & 0.4993 & 0.5354 \\
Supervised learning & 0.5142  & 0.5220  & 0.5409 & 0.5293 \\
Incremental learning   & 0.5120  & 0.5005 & 0.5032 & 0.4889 \\
Extra-device & 0.9682  & 0.5745 & 0.4959 & 0.8991 \\ \bottomrule
\end{tabular}
\end{table}

\noindent \textbf{Poisoned Fingerprint Identification.} 
{
We conduct an additional experiment to illustrate that an attacker is unable to identify the poisoned fingerprints. 
}
We consider three different attack methods:
(1) Unsupervised learning: the attacker uses clustering algorithms to divide the $(arguments, fingerprint)$ pairs into 2 clusters. 
(2) Supervised learning: the attacker randomly selects a portion of the collected $(arguments, fingerprint)$ pairs as normal ones to train models.
To identify a valid pair, the attacker predicts a fingerprint and calculates the related error with the true fingerprint.
If the error is greater than a threshold (e.g., $noise$), the attacker regards the pair poisoned. 
(3) Incremental learning: this method is almost the same as the supervised learning based method.
The difference between them lies in the way of training models.
For initialization, the attacker randomly selects a small number of collected $(arguments, fingerprint)$ pairs to train the models, then uses the models to identify the subsequent unknown pairs.
If a pair is classified as a normal (non-poisoned) one, the attacker can renew the models with this new pair.
We assume the attacker retrains the models with a fixed number
of pairs, i.e., the training step.

We test various clustering algorithms, ratios of training data and training steps for each scheme.
Also, we test different features individually.
The maximum identification accuracy for each scheme is shown in Table~\ref{tab:poison_data_identify}.
The identification of poisoned pairs is a 2-class classification task.
The highest accuracy among different schemes is around 54\%, only 4\% higher than 50\%, which indicates that software-based approaches fail to identify poisoned data.

Furthermore, we test a mixed scheme that combines software with hardware, called extra-device.
The attacker replaces models with hardware to give fingerprints.
In DAC/ADC and PWM, this scheme gets greater than 90\% accuracy, but in the other two fingerprints, the accuracy is still low. 
Accuracy is related to the discrepancy between two devices and the value of added noises.
For instance, for a $arguments$, $fp_0, fp_1$ are fingerprints from two devices.
If $|\frac{(fp_0 - fp_1)}{fp_1}| >  noise$, the poisoned pair may not be identified.
This guides us to a better way to launch poisoning, i.e., keeping $noise$ in the range of discrepancies among different devices.


\noindent \modify{\textbf{Other Parameter Settings.}}
\todo{C3}
\modify{
\ourwork prevents software mimic attacks via data poisoning and the poisoned pairs cannot be identified.
We do not experiment with other parameter settings such as $C$ in Equation~\ref{eq:poison_strategy}, as they are not key parameters and have little effect on protection effectiveness.
As long as the poisoned pairs can affect the training phases of the attackers, \ourwork works well.
The key point is the ratio of the normal pair the attackers can get and how they use it.  
These settings have been shown in the experiments above.
}

\subsubsection{\textbf{Tampering Attack}}
\begin{figure}[t]
    \centering
    \subfloat[Tampering attack success rate\label{fig:attack_tamper_sr}]{\includegraphics[width=0.23\textwidth]{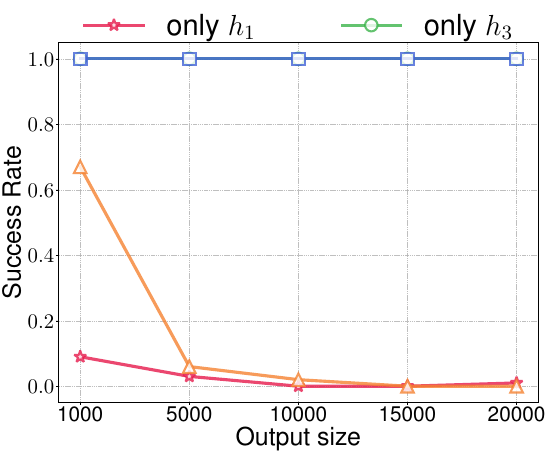}}
    \hfill 	
    \subfloat[Attack success average number\label{fig:attack_tamper_sn}]{\includegraphics[width=0.23\textwidth]{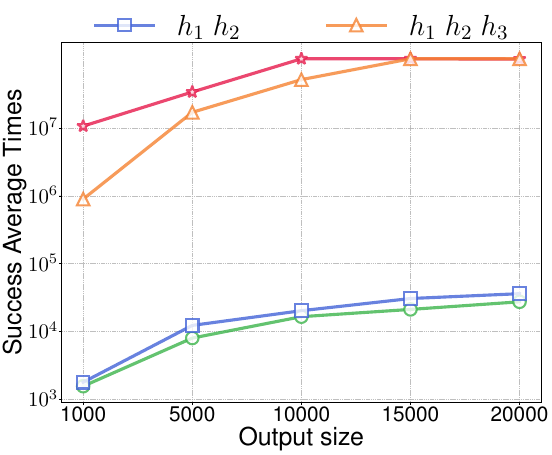}}
    \caption{Tampering attacks on \ourwork. }
    \label{fig:attack_tamper}
\end{figure}

In \ourwork, an attacker may tamper with the operation or payloads of requests.
We assume that the attacker knows the message mapping algorithm installed on the client side and the attacker tries to modify the requests and keep the tasks the same (and the tokens will be the same).
To simplify, we set parameters as below. \totalnumber is 2, the number of operation types is 200 (the car key BLE operation types in Telsa are around 40), the size of payloads is 32 bit and the size of the nonce is 16 bit.
We test the attack success rates with various numbers of arguments (i.e., the output size of the message mapping algorithm).
The results are shown in Figure~\ref{fig:attack_tamper}.  

The line whose label is "only $h_1$" means that in Algorithm~\ref{alg:mapping} we only use $h_1$ to generate arguments (the same for others).  
Figure~\ref{fig:attack_tamper_sr} shows the success rate.
Figure~\ref{fig:attack_tamper_sn} shows the average times of modifying the request to launch a successful attack.
Results prove that our algorithm is immune to tampering attacks.
With a 10,000 output space size, the attack success rate is less than 1\% and the number of arguments used in ESP32S2 is about 20,000 \modify{(i.e, the output space size is 20,000)}.
What's more, comparing the results between "only $h_1$" and "only $h_3$", we observe that it is more difficult to modify the operation than the payloads.
"$h_1\ h_2$" shows that $h_2$ raises SR as attackers can modify payloads to keep the digest the same. 
With $h_3$ the success rate reduces greatly and the success average times are almost the squared values of those without $h_3$.

\subsection{Case Studies on Various Usage Scenarios}
\label{chp:eva_overhead}
To show \ourwork's usability on different IoT devices, we choose some typical scenarios to perform case studies to evaluate the energy consumption of reasonable tasks number.

\begin{table}[t]
\caption{Power and time for fingerprint generation.}
\label{tab:overhead}
\centering
\begin{tabular}{@{}cccccc@{}}
\toprule
                                                & \textbf{Encrypt} & \textbf{Voltage} & \textbf{FPU} & \textbf{Clock} & \textbf{Storage} \\ \midrule
\multicolumn{1}{c}{\multirow{2}{*}{ESP32S2}}   & 0.23W   & 0.22W   & 0.22W       & 0.19W & 0.17W   \\
\multicolumn{1}{c}{}                           & 2ms     & 23ms    & 97ms        & 10ms  & 10ms    \\ \midrule
\multicolumn{1}{c}{\multirow{2}{*}{STM32F429}} & 0.74W   & 0.79W   & 0.76W       & 0.79W & 0.71W   \\
\multicolumn{1}{c}{}                           & 2ms     & 39ms    & 8ms         & 47ms  & 1ms     \\ \midrule
\multicolumn{1}{c}{\multirow{2}{*}{STM32F103}} & 0.15W   & 0.16W   & 0.16W       & 0.15W & 0.15W   \\
\multicolumn{1}{c}{}                           & 5ms     & 114ms   & 17ms        & 8ms   & 1ms     \\ \bottomrule
\end{tabular}
\end{table}

\subsubsection{\textbf{Smart Home}}
Smart home devices adopt trigger-action platforms~\cite{dtap18} to execute automation rules, which usually adopt token-based authentication and may be abused to maliciously trigger rules~\cite{dtap18, trigger_action2}.
We use the STM32F429 device as an IoT temperature sensor which can report the current temperature to trigger a rule of 
 "if the temperature is higher than 32$^{\circ}$C, open the window".
After adopting \ourwork, the trigger action platform can check the extra hardware access token to verify if the temperature data is actually from the sensor rather than attackers' phantom device~\cite{iot_hazards}.
Since the temperature data may be uploaded very frequently, we only use 4 fingerprints in the token to find a tradeoff between security and energy consumption.


\subsubsection{\textbf{PKE/BLE Key Fob}}
Existing PKE key fob uses rolling codes~\cite{rolling_code} for authentication and the risk is that attackers can record some codes to perform cryptographic attacks~\cite{HODOR} to reveal the generating of rolling code or reuse the rolling code.
As shown in Figure~\ref{fig:running_example}, we generate the \ourwork's access token based on the command and use the rolling code as the nonce.
We prototype the PKE rolling code mechanism on the ESP32S2 device and use two fingerprints for each request which only increases 32 bits to the existing payload.
For the BLE key fob using RSA, we can use more fingerprints (e.g., 8) as BLE can send longer payloads.
By verifying the extra access token, we can prevent cryptographic attacks~\cite{HODOR} and relay attacks on these devices.

\subsubsection{\textbf{Hardware Security Token (HST) for FIDO-U2F}}
\ourwork can be easily integrated into the existing FIDO-U2F~\cite{fido_iot} service for verifying if the FIDO-U2F HSTs are trusted devices.
We use the STM32F103 devices as a HST which implements the FIDO-U2F client and deploys the FIDO-U2F server on the PC. 
FIDO-U2F's existing counter can be used as our nonce for message mapping and generate a hardware fingerprint based token based on the response payload.
Since the HSTs have a high security requirement and are less sensitive to performance, we can generate 8 fingerprints and half of them are poisoned data.
This token can be added as extra information in the attestation certificate and the server can verify this item to check the authenticity of HSTs.
As a result, attackers attempting to clone the HST~\cite{u2f_clone} still cannot impersonate the real device, even if they have stolen the private keys.

Table~\ref{tab:overhead} shows the energy consumption when authentication with \ourwork, which varies on different fingerprinting tasks.
Compared to the baseline of default token-based authentication using AES encryption, our fingerprinting generation incur an extra energy consumption of less than 4\% on average.
The extra time consumption is less than 31ms (2 fingerprints) and 115ms (8 fingerprints) on average.


\section{Related Work}

\bsub{Hardware Fingerprints} are widely explored on various platforms such as mobile, PC, and IoT devices, to distinguish and track devices~\cite{drawnapart_gpu, clockAroundClock, SensorID} or to authenticate devices~\cite{HODOR, RFID, IoT-ID, time-print, DeMiCPU, ECU}.
Unfortunately, most of these fingerprinting features require special hardware support, such as GPU~\cite{drawnapart_gpu, clockAroundClock}, mobile sensors~\cite{SensorID, BLE/Tracking}, and NADA flash~\cite{flash-memory, time-print}, which are absent in MCU-based embedded devices.
For the approaches target IoT devices, HODOR~\cite{HODOR} and \cite{RFID} employ RFID signal features to fingerprint devices which is not a general solution for other kinds of IoT devices. 
DeMiCPU~\cite{DeMiCPU} and \cite{ECU} do not consider the attackers can MitM mimic or forward the fingerprint.
Our approach (i.e., \ourwork) aims at proposing a general fingerprint framework for all kinds of COTS MCUs that can resist MitM advisories.
IoT-ID~\cite{IoT-ID} is the closest work but they use invariant fingerprints as the device identifier, which is vulnerable to both software mimic (ML attacks) and MitM interference. 
Our work first extends IoT-ID's solution to generate variable fingerprints based on different inputs and then proposes an arguments mapping protocol to bind the inputs with specific commands to ensure the integrity of messages.


\noindent \textbf{Hardware-backed Authentication.}
Various PUF mechanisms~\cite{puf_suervey} are proposed to enforce IoT authentication by dynamically reproducing cryptographic keys from devices rather than storing the keys on the firmware which can reduce the risks of key stolen. 
Priyanka et al.~\cite{puf_suervey} investigate the performance and security of different PUF mechanisms and find that these approaches do not take both MitM attacks and software/physical impersonation attacks into consideration at the same time. 
For instance, the existing approach proposes Challenge-Response (CRP) based PUF protocols~\cite{lightweight_puf, recurrence_puf, puf_mutual} to prevent replay attacks and software impersonation attacks.  
are proposed to solve this problem. However, they cannot protect the integrity of the commands and thus are vulnerable to several MitM attacks~\cite{puf_lessons, puf_suervey}. 
Most of these PUF approaches require extra hardware supports (e.g., special circuits) which are not supported by our target devices, i.e., COTS MCU.
\ourwork aims to provide a general fingerprint framework that can be easily extended by adding new PUF-based fingerprint features  (e.g., SRAM, Flash) that do not require extra hardware support.
Moreover, we ensure the security of fingerprints by proposing argument mapping and data poison approaches to defend against MitM attacks and impersonation attacks.



\noindent \textbf{Embedded Device Authentication Security.}
Embedded device authentication has been long regarded as vulnerable~\cite{iot_hazards, attack_hw_wallet}. 
Mirai~\cite{mirai_botnet} exploits weak passwords to compromise millions of devices.
BIAS~\cite{bias} and KNOB~\cite{knob} can perform MitM attacks on almost all Bluetooth devices by impersonating validate devices.
Existing USB hardware tokens~\cite{attack_u2f, usb-mitm} are also proved to be insecure and can be cloned or impersonated during manufacturing or shipments. 
To protect these devices, various authentication or pairing approaches are proposed.  
T2Pair~\cite{t2pair}, \cite{contauth}, and \cite{wearable_pair} utilize the sensing operations (e.g., knob, button, or touch screen) to secure the pair of IoT devices.
Their limitation is requiring Human-in-the-loop to generate sensor data.
Using hardware features (e.g., fingerprint~\cite{DeMiCPU, time-print}, PUF~\cite{multi-factor-puf, lightweight_multi_factor}) to secure the authentication is widely discussed in various platforms.
However, IoT-ID~\cite{IoT-ID} is the only work that focuses on MCU-based devices.
We address several applicable issues of IoT-ID by proposing a new hardware fingerprint authentication framework \ourwork, which can work on all kinds of MCU devices and can resist traditional token compromise and MitM attacks.

\section{Discussion}
\label{sec:discussion}



\bsub{\modify{Attackers Compromise the Devices.}}
\todo{A1}
\modify{
\ourwork aims to mitigate device impersonation attacks due to credential theft, weak cryptographic support, or insecure authentication implementation.
For attackers who have compromised the system locally or remotely, they may manipulate this device to send requests with valid authentication data (e.g., \ourwork's hardware fingerprints), which is outside the scope of authentication security and not the design goal of \ourwork.
If they want to clone~\cite{u2f_clone, clone_key} this device for off-path exploitation (e.g., via Phantom Client~\cite{iot_hazards}), they still need to collect enough fingerprint data to mimic the real hardware features.
However, the specific fingerprinting parameters used for data collection are unknown to attackers. 
Therefore, attackers need to explore a large number of fingerprinting parameters to obtain the set of training data, which is time-consuming and can be easily detected by the backends or the device owner.
}

\bsub{The Maximum Limit of Requests Supported.}
In \ourwork, the maximum number of requests that can be issued is determined by the range of hardware task arguments. In fact, the argument range for hardware tasks is typically large enough to accommodate the number of device requests in practical scenarios. 
\todo{C4}
\modify{
For example, in a PKE/BLE key fob scenario, there are 20,000 different argument values on the ESP32S2 device.
Assuming a person sends five distinct requests per day (e.g. unlocking the door) and 4 fingerprints are used per authentication, \ourwork can provide protection for approximately 1,000 days.
In addition, we can extend the current lifetime of \ourwork by extracting new fingerprints from existing hardware features and retraining the backend.
For instance, changing the SRAM address ranges can generate different fingerprints.
All fingerprinting tasks and arguments can be changed to get more fingerprints for new requests.
}

\bsub{Device Aging.} 
In practice, hardware fingerprints may change due to device damage, aging, or other factors, resulting in failed server authentication at the backend side.
Regarding this issue, customers can securely return their devices to the backend for re-collection of fingerprint data. Thus, these devices can be successfully authenticated when they are re-deployed in the customers' environments.


\bsub{\modify{\ourwork with Fewer Hardware Features.}}
\todo{B3}
\modify{
Hardware features used in \ourwork may not be available on all devices, such as the DAC.
But \ourwork can still work, because fewer hardware features do not mean fewer fingerprints.
On the one hand, SRAM and RTCs are available on almost all types of MCUs. Also, SRAM and RTCs can provide sufficient fingerprints.
On the other hand, we can increase the number of fingerprints by modifying the fingerprinting tasks to generate as many fingerprints as possible.
}

\noindent \textbf{Future Works.}
\ourwork provides a general hardware fingerprinting scheme.
It is promising to extend \ourwork by supporting more hardware features and PUFs~\cite{puf-taxonomy}.


\section{Conclusion}
We introduce \ourwork, a hardware fingerprint based authentication mechanism to enhance the security of existing token-based authentication approaches for MCU-based IoT devices.
\ourwork can protect device authentication when traditional cryptography-based approaches (e.g., message encryption and signature) are compromised by attackers.
With its simplicity, \ourwork can be applied in diverse scenarios to authenticate different MCU-based IoT devices with high accuracy and can resist common attacks. 
The \ourwork solution can be easily integrated into all kinds of existing IoT devices as its client runtime supports major COTS MCUs and its backend authentication service can be deployed on common IoT devices or on the cloud.

\section*{Acknowledgment}

We would like to thank our shepherd and the anonymous reviewers for their comments.
This work is supported in part by the Natural Science Foundation of China under Grant U20B2049, U21B2018, 62302452, and 62132011; Zhejiang Provincial Natural Science Foundation of China under Grant LQ23F020019.
Kun Sun's work is supported in part by National Centers of Academic Excellence in Cybersecurity under Grant \#H98230-22-1-0311.



\loadtail{main}
\loadappend{}

\section{Details of Designed Tasks}
\label{chp:hardware_fingerprints_details}
The following are the designed details of the tasks for 6 features, along with their arguments and corresponding outputs (i.e., fingerprints):

\noindent \textbf{DAC/ADC}. We use DAC to convert a number to an analog voltage and use ADC to read it, then calculate the error between the read voltage and the theory voltage as the output. The arguments are, (1) the value of DAC input; (2) the working state of voltage drain drain; (3) the format of ADC voltage, the raw value or the corrected value; (4) the output mode, including different error representations and different pins of ADC.

\noindent \textbf{FPU}. We use the calculation of Mandelbrot fractal calculating, which is a way to test the speed of FPU. The arguments are, (1) whether FPU is used; (2) the x-bound of Mandelbrot set; (3) the y-bound of Mandelbrot set. The output is time spent of calculating.

\noindent \textbf{PWM}. We utilize ADC to measure voltages generated by PWM, and the output is calculated as the sum of voltages over multiple periods. The arguments are, (1) the clock source of PWM; (2) the frequency of the clock; (3) the number of measured period; (4) the working state of voltage drain drain; (5) the duty ratio of PWM.

\noindent \textbf{RTCFre and RTCPha}.During frequency testing, we measure the time it takes to complete several periods as the output. The arguments are: (1) the clock source; (2) the number of clock division; (3) the adjusting value, which is used to adjust clock during different environments; (4) the number of measured period. Also, we measure the instantaneous phase of source clock. The arguments are: (1) the clock source; (2) the number of clock divisions; (3) the supposed period of clock ticking.

\noindent \textbf{SRAM}. The input is a target address (required to be 4-aligned) of SRAM as the start address. And we make the following 32-bit (contains the start address) into an integer as output.

\section{\modify{Security Proof of \ourwork}}

\todo{C1}
\modify{
\ourwork relies on data poisoning to defend against software mimic attacks and uses the message mapping algorithm to defend against tampering attacks.
In this section, we provide theoretical proofs of the security of these two key designs.
Combined with the analysis in \S~\ref{chp:sec_analysis}, we can demonstrate the security of the entire system of \ourwork.
}

\subsection{\modify{Proof of Data Poisoning Effectiveness}}
\label{chp:proof_poison}


\noindent \modify{
\textbf{Influence of Poisoned Data on Model Learning.} We formulate how the poisoned data can affect the model to learn a linear mapping as a regression problem and solve it:
}

\textit{Problem statement:} There is mapping $Y=a*X+b$, and pairs in the mapping can be formatted as $(x,y)$. Now, we transform all $(x,y)$ to $(x,y'), y'=cy+d$. If an attacker learns with the modified pairs and aims to make the mean square error as small as possible, what mapping will be learned?

\textit{Solution:} The mapping learned by the attacker is still a linear one, which can be formatted as $Y'=a'*X+b'$. According to the least square method, we can calculate $a',b'$.
\begin{equation}
\begin{aligned}
a'&=\frac{\Sigma (x - \overline{x})(y'-\overline{y'})}{\Sigma (x - \overline{x})^2} \\
&=\frac{\Sigma (x - \overline{x})(cy+d-c\overline{y} - d)}{\Sigma (x - \overline{x})^2} \\
&=ca \\
b' &= \overline{y'} - a'\overline{x} \\
&=c\overline{y}- ca\overline{x} +d  \\
&=cb+d
\end{aligned}
\end{equation}

\noindent \modify{
\textbf{The Effectiveness of \ourwork's Poisoned Fingerprints Identification.} 
We show how \ourwork identifies the poisoned fingerprints based on the solution.
}
\modify{
The backend checks whether a fingerprint is legal or not by comparing it to the raw fingerprint and can tolerate the hardware bias.
If a fingerprint is within the bias, the backend accepts it, otherwise the fingerprint is rejected.
We use $y$ for the fingerprint and $x$ for the arguments.
According to the proof, when it comes to a new $x$, an attacker will give a $y'$.
Compared to the original $y$ we have $\Delta = |\frac{y-y'}{y}| = (c-1)+|\frac{d}{ax+b}|$.
As long as $\Delta$ is greater than the hardware bias, $y'$ will not match the original $y$ and will be rejected by the backend.
}

\subsection{\modify{Proof of Message Mapping Algorithm' Security}}
\label{chp:proof_tamper}


\noindent \modify{\textbf{Proof of the Hash Collision Security.}
To analyze the security of Algorithm~\ref{alg:mapping}, we construct and solve a hash collision problem and prove it can ensure collision security under our settings.
} 

\textit{Problem statement:} There is hash function $H$ whose output space size is $d$. Payloads are $p_0,p_1$. $c$ is a command. The results are $h_0=H(c||p_0||p_1), h_1=H(c||p_1||p_0)$. The purpose of the attacker is to modify $c$ to $c'$ and to keep $h_0'= h_0, h_1'=h_1$. Solve the probability.

\textit{Solution:} The $c'$ is fixed and the attacker modifies $p_0,p_1$. For a fixed $p_0'$, as the output of $H$ is uniformed,
\begin{equation}
P\{h_0=h_0'|p_0',p_1'\} = \frac{1}{d}
\end{equation}
Although in $h_0.h_1$ the difference is only the position of $p_0,p_1$, the outputs will be completely different and also be seen as uniform.
\begin{equation}
P\{h_0=h_0', h_1=h_1'|p_0',p_1'\} = \frac{1}{d^2}
\end{equation}
We assume there are $n$ kinds of combinations for $p_0,p_1$, Then
\begin{equation}
P\{h_0=h_0', h_1=h_1'\} = 1-(\frac{d^2-1}{d^2})^n
\end{equation}

\noindent \modify{\textbf{A Step-by-step Analysis of Algorithm~\ref{alg:mapping}.}}
\modify{
Based on this proof, we explain the security of the whole message mapping algorithm.
We use the case where there are only two tasks, i.e. the algorithm produces 2 tasks, each of which has $d_{0}$ different possible values (we use "space size" to represent the number of possible values).
The attacker aims to tamper with the request and keep the output tasks the same.
}

\modify{
\textbf{With only a single hash function}: if the generation of two tasks is independent, the attacker only needs to tamper with tasks whose space size is $d_0$ twice.
In total, the attacker's attempt times are at most $2d_{0}$.
}

\modify{
\textbf{With \bm{$h_1$} (line 8)}: $h_1$ links the generation of the two tasks.
If the attacker replaces the operation or the nonce, both tasks will change, which means that the attacker needs to consider the space size of the combination of two tasks.
The space size of the combination of two tasks is $d_{0}*d_{0}$, i.e. the attempt times are at most $d_{0}^2$.
}

\modify{
\textbf{With \bm{$h_1, h_2$} (line 10)}: $h_2$ provides integrity protection for payloads, but allows the attacker to modify payloads to keep the same tasks.
The attacker can modify $h_2$ respectively for the two tasks, and the attempt times are at most $2d_{0}$.
}

\modify{
\textbf{With \bm{$h_1, h_2, h_3$} (line 12}): To solve the problem posed by $h_2$, we add $h_3$.
According to the proof, with $h_3$ the attacker's attempt times are at most $d_{0}^2$.
Furthermore, if there are more generations of tasks linked together, the difficulty for attackers to manipulate the request will increase.
}

\modify{
In \ourwork, the output space of ESP32S2 devices is 20,000 (i.e. $d$), and the success rate for an attacker is about 2\% (i.e. $P$) with $10^7$ (i.e. $n$) attempt times.
}

\section{Performance of Different Model Combination}
\label{sec:model}

We compare the effectiveness of different models in predictors and verifiers of \ourwork backend's authentication service.
These models are tested on the DAC/ADC features of ESP32S2 devices. 
The TPRs and FPRs are shown in Table~\ref{tab:eva_model}. The predictor models are in the rows and the verifier models are in the columns. The results show that different models have little effect on \ourwork. 

\begin{table}[t]
\caption{Different combinations of models.}
\centering
\label{tab:eva_model}
\begin{tabular}{@{}cccc@{}}
\toprule
             & RandomForest & ExtraTree & DecisionTree \\ \midrule
RandomForest &   0.85,0,09  &  0.83,0.08  &  0.83,0.08  \\
ExtraTree    &   0.85,0.09  &  0.84,0.08  &  0.84,0.08   \\
DecisionTree &   0.85,0,09  &  0.83,0.08  &  0.84,0.08   \\ \bottomrule
\end{tabular}
\end{table}

\balance

\clearpage

\nobalance
\section{Artifact Appendix}

This artifact contains the source code of \ourwork and the instructions to run it.
\ourwork is designed for authenticating embedded devices via hardware fingerprinting. 
To evaluate the basic functionality, you need at least one of these development boards: \ESPClickBuy, \STMFFourClickBuy, or \STMFOneClickBuy.
If you do not have any of these devices, we offer a demo that can be executed on the \ReNodeClick emulator to showcase the functionality.
Furthermore, we provide a dataset obtained from our physical devices, allowing you to reproduce the paper's experimental results without necessitating any IoT hardware.

\subsection{Description \& Requirements}




\subsubsection{How to access}
All the documents and source code are available on github: \url{https://github.com/IoTAccessControl/MCU-Token/tree/master}. 
And the DOI link is \url{https://zenodo.org/doi/10.5281/zenodo.10117167}. 

\subsubsection{Hardware dependencies} \ourwork is implemented on the following devices, ESP32S2, STM32F429, and STM32F103. Make sure you have at least one of them to collect fingerprint data and validate the results.

\subsubsection{Software dependencies} In summary, to compile the source code and deploy \ourwork, you need to install one of the following software.
\begin{itemize}
    \item ESP32-idf for ESP32S2
    \item Keil for STM32F429 and STM32F103
    \item Renode for emulation
\end{itemize}

\subsubsection{Benchmarks} None.

\subsection{Artifact Installation \& Configuration}


\begin{enumerate}
    \item Install Python(3.8) and other required software.
    \item Clone the source code from \href{https://github.com/IoTAccessControl/MCU-Token/tree/master}{the repo}.
    \item Install the requirements from \href{https://github.com/IoTAccessControl/MCU-Token/tree/master/MCUToken/server}{requirements.txt}.
\end{enumerate}

\subsection{Experiment Workflow}



The detailed steps for generating fingerprints for a device and evaluating them are as follows:

\label{install_ourwork}
1. Install \ourwork on your devices and ensure that the wires are connected correctly (according to \href{https://github.com/IoTAccessControl/MCU-Token/blob/master/Device-porting/readme.md}{Device-porting/README.md}). 

2. Collect training data through the serial port. We provide shell scripts to collect fingerprint data (see \href{https://github.com/IoTAccessControl/MCU-Token/tree/master/MCUToken/server}{MCU-Token/server}). For example, 
\begin{lstlisting}
    python bat_generator.py
        --device_number 0 --port COM3
\end{lstlisting}

\label{log_generate}
3. Evaluate the accuracy of fingerprint verification. We provide shell scripts for generating logs which contain the core results for generating the figures in our paper (see \href{https://github.com/IoTAccessControl/MCU-Token/tree/master/MCUToken/server#reproduce-results-in-the-paper}{reproducable}). Such as, 
\begin{lstlisting}
    bash 0_generate_ref_log.sh
\end{lstlisting}

\subsection{Major Claims}



\begin{itemize}
    \item (C1): We can generate a hardware-based access token for each command and collect fingerprints for each device. This is proven by the experiment (E1).
    \item (C2): We evaluate the performance of the tokens (hardware fingerprints) with different settings. Including the accuracy of authentication, the robustness in different environments and the effectiveness of defense against three types of attacks. This is proven by the experiment (E2).
\end{itemize}

\subsection{Evaluation}





\subsubsection{Experiment (E1)} [30 human-minutes]: Generate hardware-based access tokens for commands and extract hardware fingerprints for devices. Details are shown in \href{https://github.com/IoTAccessControl/MCU-Token/blob/master/Device-porting/readme.md}{Device-porting/README.md}.

\textit{[Preparation]} Install \ourwork on your device or open Renode. If you are using a physical device, make sure the wire connection is correct.

\textit{[Execution]} For a physical device, open the serial port and use the "token\_gen" command to generate a token for the command. For example,
\begin{lstlisting}
    token_gen SET_TEM SEAT1 25
\end{lstlisting}
Use the "fp\_gen" to extract hardware fingerprints, for example,
\begin{lstlisting}
    fp_gen STM32 11010 0 0
\end{lstlisting}
If you do not have a physical device, you can follow the steps in the document to use Renode.

\textit{[Results]} If you use "token\_gen", the command token is printed to the serial port. If you use "fp\_gen", the results of fingerprint tasks are printed to the serial port. In a physical device, you will see u8 serials (unreadable). And in Renode you will get readable results (strings).

\subsubsection{Experiment (E2)} [30 human-minutes + 2 compute-hours]: Evaluate the performance of the hardware tokens (fingerprints). The details are shown in document \href{https://github.com/IoTAccessControl/MCU-Token/tree/master/MCUToken/server#reproduce-results-in-the-paper}{Reproducable}.

\textbf{TL;DR}~Run Step-2 and get the results in our paper.

\textit{[Preparation]} Install Python(3.8). 

\textit{[Execution]} Step-1: You can generate the evaluation results with the provided "*\_log.sh" scripts (may take several hours). 
After replacing the \href{https://github.com/IoTAccessControl/MCU-Token/tree/master/MCUToken/server/evaluation/result}{original results} with yours, you can use the plotting programs to get the figures and tables that are similar to or the same as those in the paper. You can train your own models based on the dataset provided by us and evaluate the data of your devices. Step-2: You can run the plotting programs to get the figures and tables based on our data, for example, 
\begin{lstlisting}
    python3 Fig-2_plot.py
\end{lstlisting}

\textit{[Results]} With Step-1, you can get the raw results logs and get the evaluation results. With Step-2, you can reproduce all the tables and figures in our paper.



\end{document}